%\input psfig
%%%%%%%%%%%%%%%%%%%%%%%%%%%%%%%%%%%%%%%
%%%%%%%%%%%%%%%%%%  tex macros for preprints, cm version %%%%%%%%%%%%%%
%         (P. Ginsparg <ginsparg@lanl.gov>, last updated 7/94)
%                if confused, type `b' in response to query 
%           hypertex extensions (still provisional), 7/26/94
%
%---------------------------------------------------------------------%
%\input hyperbasics %comment out this line to restore non-hyper functionality
%
%% site dependent options:
%% \unredoffs and \redoffs define horizontal and vertical offsets
%% respectively for unreduced and reduced modes. \speclscape defines
%% the \special{} call that sets printer to landscape (sideways) mode.
%% from standard set below, leave uncommented as appropriate or redefine
%
%%% next 400dpi
\def\unredoffs{} \def\redoffs{\voffset=-.31truein\hoffset=-.48truein}
\def\speclscape{}
%\def\speclscape{\special{papersize=11in,8.5in}}
%
%%% apple lw
%\def\unredoffs{} \def\redoffs{\voffset=-.31truein\hoffset=-.59truein}
%\def\speclscape{\special{ps: landscape}}
%
%%% qms lasergrafix:
%\def\unredoffs{} \def\redoffs{\voffset=-.4truein\hoffset=.125truein}
%\def\speclscape{\special{qms: landscape}}
%
%%% saclay A4 paper:
%\def\unredoffs{\hoffset-.14truein\voffset-.2truein}
%\def\redoffs{\voffset=-.45truein\hoffset=-.21truein}
%\def\speclscape{\special{landscape}}
%
%---------------------------------------------------------------------%
%
\newbox\leftpage \newdimen\fullhsize \newdimen\hstitle \newdimen\hsbody
\tolerance=1000\hfuzz=2pt
\catcode`\@=11 % This allows us to modify PLAIN macros.
\ifx\hyperdef\UNd@FiNeD\def\hyperdef#1#2#3#4{#4}\def\hyperref#1#2#3#4{#4}\fi
\def\bigans{b }
\def\answ{b }
%\message{ big or little (b/l)? }\read-1 to\answ
%
\ifx\answ\bigans\message{(This will come out unreduced.}
\magnification=1200\unredoffs\baselineskip=16pt plus 2pt minus 1pt
\hsbody=\hsize \hstitle=\hsize %take default values for unreduced format
\else\message{(This will be reduced.} \let\l@r=L
\magnification=1000\baselineskip=16pt plus 2pt minus 1pt \vsize=7truein
\redoffs \hstitle=8truein\hsbody=4.75truein\fullhsize=10truein\hsize=\hsbody
\output={\ifnum\pageno=0 %%% This is the HUTP version
  \shipout\vbox{\speclscape{\hsize\fullhsize\makeheadline}
    \hbox to \fullhsize{\hfill\pagebody\hfill}}\advancepageno
  \else
  \almostshipout{\leftline{\vbox{\pagebody\makefootline}}}\advancepageno
  \fi}
\def\almostshipout#1{\if L\l@r \count1=1 \message{[\the\count0.\the\count1]}
      \global\setbox\leftpage=#1 \global\let\l@r=R
 \else \count1=2
  \shipout\vbox{\speclscape{\hsize\fullhsize\makeheadline}
      \hbox to\fullhsize{\box\leftpage\hfil#1}}  \global\let\l@r=L\fi}
\fi
%---------------------------------------------------------------------
%
\newcount\yearltd\yearltd=\year\advance\yearltd by -2000

\def\Title#1#2{\nopagenumbers\abstractfont\hsize=\hstitle\rightline{#1}%
\vskip 1in\centerline{\titlefont #2}\abstractfont\vskip .5in\pageno=0}
\def\Date#1{\vfill\leftline{#1}\tenpoint\supereject\global\hsize=\hsbody%
\footline={\hss\tenrm\hyperdef\hypernoname{page}\folio\folio\hss}}%
% (restores pagenumbers)
%
%       use following instead of \Date on the preliminary draft,
%       puts date/time on each page in big mode, writes labels in margins

\def\draftmode{\message{ DRAFTMODE }\def\draftdate{{\rm preliminary draft:
\number\month/\number\day/\number\yearltd\ \ \hourmin}}%
\headline={\hfil\draftdate}\writelabels\baselineskip=20pt plus 2pt minus 2pt
 {\count255=\time\divide\count255 by 60 \xdef\hourmin{\number\count255}
  \multiply\count255 by-60\advance\count255 by\time
  \xdef\hourmin{\hourmin:\ifnum\count255<10 0\fi\the\count255}}}
%       use \nolabels to get rid of eqn, ref, and fig labels in draft mode
\def\nolabels{\def\wrlabeL##1{}\def\eqlabeL##1{}\def\reflabeL##1{}}
\def\writelabels{\def\wrlabeL##1{\leavevmode\vadjust{\rlap{\smash%
{\line{{\escapechar=` \hfill\rlap{\sevenrm\hskip.03in\string##1}}}}}}}%
\def\eqlabeL##1{{\escapechar-1\rlap{\sevenrm\hskip.05in\string##1}}}%
\def\reflabeL##1{\noexpand\llap{\noexpand\sevenrm\string\string\string##1}}}
\nolabels
%
% tagged sec numbers
\global\newcount\secno \global\secno=0
\global\newcount\meqno \global\meqno=1
\def\s@csym{}
\def\newsec#1{\global\advance\secno by1%
{\toks0{#1}\message{(\the\secno. \the\toks0)}}%
%\ifx\answ\bigans \vfill\eject \else \bigbreak\bigskip \fi  %if desired
\global\subsecno=0\eqnres@t\let\s@csym\secsym\xdef\secn@m{\the\secno}\noindent
{\bf\hyperdef\hypernoname{section}{\the\secno}{\the\secno.} #1}%
\writetoca{{\string\hyperref{}{section}{\the\secno}{\it\the\secno.}} {{\it #1} }}%
\par\nobreak\medskip\nobreak}
\def\eqnres@t{\xdef\secsym{\the\secno.}\global\meqno=1\bigbreak\bigskip}
\def\sequentialequations{\def\eqnres@t{\bigbreak}}\xdef\secsym{}
\global\newcount\subsecno \global\subsecno=0
\def\subsec#1{\global\advance\subsecno by1%
{\toks0{#1}\message{(\s@csym\the\subsecno. \the\toks0)}}%
\ifnum\lastpenalty>9000\else\bigbreak\fi       \global\subsubsecno=0
\noindent{\it\hyperdef\hypernoname{subsection}{\secn@m.\the\subsecno}%
{\secn@m.\the\subsecno.} #1}\writetoca{\string\quad
{\string\hyperref{}{subsection}{\secn@m.\the\subsecno}{\secn@m.\the\subsecno.}}
{#1}}\par\nobreak\medskip\nobreak}
\def\appendix#1#2{\global\meqno=1\global\subsecno=0\xdef\secsym{\hbox{#1.}}%
\bigbreak\bigskip\noindent{\bf Appendix \hyperdef\hypernoname{appendix}{#1}%
{#1.} #2}{\toks0{(#1. #2)}\message{\the\toks0}}%
\xdef\s@csym{#1.}\xdef\secn@m{#1}%
\writetoca{\string\hyperref{}{appendix}{#1}{{\it Appendix} {\it #1.}} {\it #2}}%
\par\nobreak\medskip\nobreak}
%
%       \eqn\label{a+b=c}	gives displayed equation, numbered
%				consecutively within sections.
%     \eqnn and \eqna define labels in advance (of eqalign?)
%
\def\checkm@de#1#2{\ifmmode{\def\f@rst##1{##1}\hyperdef\hypernoname{equation}%
{#1}{#2}}\else\hyperref{}{equation}{#1}{#2}\fi}
\def\eqnn#1{\DefWarn#1\xdef #1{(\noexpand\relax\noexpand\checkm@de%
{\s@csym\the\meqno}{\secsym\the\meqno})}%
\wrlabeL#1\writedef{#1\leftbracket#1}\global\advance\meqno by1}
\def\f@rst#1{\c@t#1a\em@ark}\def\c@t#1#2\em@ark{#1}
\def\eqna#1{\DefWarn#1\wrlabeL{#1$\{\}$}%
\xdef #1##1{(\noexpand\relax\noexpand\checkm@de%
{\s@csym\the\meqno\noexpand\f@rst{##1}}{\hbox{$\secsym\the\meqno##1$}})}
\writedef{#1\numbersign1\leftbracket#1{\numbersign1}}\global\advance\meqno by1}
\def\eqn#1#2{\DefWarn#1%
\xdef #1{(\noexpand\hyperref{}{equation}{\s@csym\the\meqno}%
{\secsym\the\meqno})}$$#2\eqno(\hyperdef\hypernoname{equation}%
{\s@csym\the\meqno}{\secsym\the\meqno})\eqlabeL#1$$%
\writedef{#1\leftbracket#1}\global\advance\meqno by1}
\def\xeqn{\expandafter\xe@n}\def\xe@n(#1){#1}
\def\xeqna#1{\expandafter\xe@n#1}
\def\eqns#1{(\e@ns #1{\hbox{}})}
\def\e@ns#1{\ifx\UNd@FiNeD#1\message{eqnlabel \string#1 is undefined.}%
\xdef#1{(?.?)}\fi{\let\hyperref=\relax\xdef\next{#1}}%
\ifx\next\em@rk\def\next{}\else%
\ifx\next#1\xeqn#1\else\def\n@xt{#1}\ifx\n@xt\next#1\else\xeqna#1\fi
\fi\let\next=\e@ns\fi\next}

\def\DefWarn#1{\ifx\UNd@FiNeD#1\else
\immediate\write16{*** WARNING: the label \string#1 is already defined ***}\fi}
%
%			 footnotes
\newskip\footskip\footskip14pt plus 1pt minus 1pt %sets footnote baselineskip
\def\footnotefont{\ninepoint}\def\f@t#1{\footnotefont #1\@foot}
\def\f@@t{\baselineskip\footskip\bgroup\footnotefont\aftergroup\@foot\let\next}
\setbox\strutbox=\hbox{\vrule height9.5pt depth4.5pt width0pt}
\global\newcount\ftno \global\ftno=0
\def\foot{\global\advance\ftno by1\def\foot@rg{\hyperref{}{footnote}%
{\the\ftno}{\the\ftno}\xdef\foot@rg{\noexpand\hyperdef\noexpand\hypernoname%
{footnote}{\the\ftno}{\the\ftno}}}\footnote{$^{\foot@rg}$}}
%
%say \footend to put footnotes at end
%will cause problems if \ref used inside \foot, instead use \nref before
\newwrite\ftfile
\def\footend{\def\foot{\global\advance\ftno by1\chardef\wfile=\ftfile
%%$^{\the\ftno}$\ifnum\ftno=1\immediate\openout\ftfile=\jobname.fts\fi%
\hyperref{}{footnote}{\the\ftno}{$^{\the\ftno}$}%
\ifnum\ftno=1\immediate\openout\ftfile=\jobname.fts\fi%
\immediate\write\ftfile{\noexpand\smallskip%
%%\noexpand\item{f\the\ftno:\ }\pctsign}\findarg}%
\noexpand\item{\noexpand\hyperdef\noexpand\hypernoname{footnote}
{\the\ftno}{f\the\ftno}:\ }\pctsign}\findarg}%
\def\footatend{\vfill\eject\immediate\closeout\ftfile{\parindent=20pt
\centerline{\bf Footnotes}\nobreak\bigskip\input \jobname.fts }}}
\def\footatend{}
%
%     \ref\label{text}
% generates a number, assigns it to \label, generates an entry.
% To list the refs on a separate page,  \listrefs
%
\global\newcount\refno \global\refno=1
\newwrite\rfile
\def\ref{[\hyperref{}{reference}{\the\refno}{\the\refno}]\nref}
\def\nref#1{\DefWarn#1%
\xdef#1{[\noexpand\hyperref{}{reference}{\the\refno}{\the\refno}]}%
\writedef{#1\leftbracket#1}%
\ifnum\refno=1\immediate\openout\rfile=\jobname.refs\fi
\chardef\wfile=\rfile\immediate\write\rfile{\noexpand\item{[\noexpand\hyperdef%
\noexpand\hypernoname{reference}{\the\refno}{\the\refno}]\ }%
\reflabeL{#1\hskip.31in}\pctsign}\global\advance\refno by1\findarg}
%	horrible hack to sidestep tex \write limitation
\def\findarg#1#{\begingroup\obeylines\newlinechar=`\^^M\pass@rg}
{\obeylines\gdef\pass@rg#1{\writ@line\relax #1^^M\hbox{}^^M}%
\gdef\writ@line#1^^M{\expandafter\toks0\expandafter{\striprel@x #1}%
\edef\next{\the\toks0}\ifx\next\em@rk\let\next=\endgroup\else\ifx\next\empty%
\else\immediate\write\wfile{\the\toks0}\fi\let\next=\writ@line\fi\next\relax}}
\def\striprel@x#1{} \def\em@rk{\hbox{}}
\def\lref{\begingroup\obeylines\lr@f}
\def\lr@f#1#2{\DefWarn#1\gdef#1{\let#1=\UNd@FiNeD\ref#1{#2}}\endgroup\unskip}

\def\addref#1{\immediate\write\rfile{\noexpand\item{}#1}} %now unnecessary
\def\listrefs{\footatend\vfill\supereject\immediate\closeout\rfile\writestoppt
\baselineskip=\footskip\centerline{{\bf References}}\bigskip{\parindent=20pt%
\frenchspacing\escapechar=` \input \jobname.refs\vfill\eject}\nonfrenchspacing}
\def\startrefs#1{\immediate\openout\rfile=\jobname.refs\refno=#1}
\def\xref{\expandafter\xr@f}\def\xr@f[#1]{#1}
\def\refs#1{\count255=1[\r@fs #1{\hbox{}}]}
\def\r@fs#1{\ifx\UNd@FiNeD#1\message{reflabel \string#1 is undefined.}%
\nref#1{need to supply reference \string#1.}\fi%
\vphantom{\hphantom{#1}}{\let\hyperref=\relax\xdef\next{#1}}%
\ifx\next\em@rk\def\next{}%
\else\ifx\next#1\ifodd\count255\relax\xref#1\count255=0\fi%
\else#1\count255=1\fi\let\next=\r@fs\fi\next}
%

%
% this is ugly, but moore insists
\newwrite\ffile\global\newcount\figno \global\figno=1
\def\fig{fig.~\hyperref{}{figure}{\the\figno}{\the\figno}\nfig}
\def\nfig#1{\DefWarn#1%
\xdef#1{fig.~\noexpand\hyperref{}{figure}{\the\figno}{\the\figno}}%
\writedef{#1\leftbracket fig.\noexpand~\xfig#1}%
\ifnum\figno=1\immediate\openout\ffile=\jobname.figs\fi\chardef\wfile=\ffile%
{\let\hyperref=\relax
\immediate\write\ffile{\noexpand\medskip\noexpand\item{Fig.\ %
\noexpand\hyperdef\noexpand\hypernoname{figure}{\the\figno}{\the\figno}. }
\reflabeL{#1\hskip.55in}\pctsign}}\global\advance\figno by1\findarg}
\def\listfigs{\vfill\eject\immediate\closeout\ffile{\parindent40pt
\baselineskip14pt\centerline{{\bf Figure Captions}}\nobreak\medskip
\escapechar=` \input \jobname.figs\vfill\eject}}
\def\xfig{\expandafter\xf@g}\def\xf@g fig.\penalty\@M\ {}
\def\figs#1{figs.~\f@gs #1{\hbox{}}}
\def\f@gs#1{{\let\hyperref=\relax\xdef\next{#1}}\ifx\next\em@rk\def\next{}\else
\ifx\next#1\xfig #1\else#1\fi\let\next=\f@gs\fi\next}
\def\figin{\epsfcheck\figin}\def\figins{\epsfcheck\figins}
\def\epsfcheck{\ifx\epsfbox\UNd@FiNeD
\message{(NO epsf.tex, FIGURES WILL BE IGNORED)}
\gdef\figin##1{\vskip2in}\gdef\figins##1{\hskip.5in}% blank space instead
\else\message{(FIGURES WILL BE INCLUDED)}%
\gdef\figin##1{##1}\gdef\figins##1{##1}\fi}
\def\DefWarn#1{}
\def\figinsert{\goodbreak\midinsert}
\def\ifig#1#2#3{\DefWarn#1\xdef#1{Fig.~\noexpand\hyperref{}{figure}%
{\the\figno}{\the\figno}}\writedef{#1\leftbracket fig.\noexpand~\xfig#1}%
\figinsert\figin{\centerline{#3}}\medskip\centerline{\vbox{\baselineskip12pt
\advance\hsize by -1truein\noindent\wrlabeL{#1=#1}\footnotefont%
{\bf Fig.~\hyperdef\hypernoname{figure}{\the\figno}{\the\figno}:} #2}}
\bigskip\endinsert\global\advance\figno by1}
\newwrite\lfile
{\escapechar-1\xdef\pctsign{\string\%}\xdef\leftbracket{\string\{}
\xdef\rightbracket{\string\}}\xdef\numbersign{\string\#}}
\def\writedefs{\immediate\openout\lfile=\jobname.defs \def\writedef##1{%
{\let\hyperref=\relax\let\hyperdef=\relax\let\hypernoname=\relax
 \immediate\write\lfile{\string\def\string##1\rightbracket}}}}%
\def\writestop{\def\writestoppt{\immediate\write\lfile{\string\pageno
 \the\pageno\string\startrefs\leftbracket\the\refno\rightbracket
 \string\def\string\secsym\leftbracket\secsym\rightbracket
 \string\secno\the\secno\string\meqno\the\meqno}\immediate\closeout\lfile}}
\def\writestoppt{}\def\writedef#1{}
\def\seclab#1{\DefWarn#1%
\xdef #1{\noexpand\hyperref{}{section}{\the\secno}{\the\secno}}%
\writedef{#1\leftbracket#1}\wrlabeL{#1=#1}}
\def\subseclab#1{\DefWarn#1%
\xdef #1{\noexpand\hyperref{}{subsection}{\secn@m.\the\subsecno}%
{\secn@m.\the\subsecno}}\writedef{#1\leftbracket#1}\wrlabeL{#1=#1}}
\def\applab#1{\DefWarn#1%
\xdef #1{\noexpand\hyperref{}{appendix}{\secn@m}{\secn@m}}%
\writedef{#1\leftbracket#1}\wrlabeL{#1=#1}}
\newwrite\tfile \def\writetoca#1{}
\def\leaderfill{\leaders\hbox to 1em{\hss.\hss}\hfill}
%	use this to write file with table of contents
\def\writetoc{\immediate\openout\tfile=\jobname.toc
   \def\writetoca##1{{\edef\next{\write\tfile{\noindent ##1
   \string\leaderfill {\string\hyperref{}{page}{\noexpand\number\pageno}%
                       {\noexpand\number\pageno}} \par}}\next}}}
%       and this lists table of contents on second pass
\newread\ch@ckfile
\def\listtoc{\immediate\closeout\tfile\immediate\openin\ch@ckfile=\jobname.toc
\ifeof\ch@ckfile\message{no file \jobname.toc, no table of contents this pass}%
\else\closein\ch@ckfile\centerline{\bf Contents}\nobreak\medskip%
{\baselineskip=18.5pt  \footnotefont
\parskip=2pt\catcode`\@=12\input\jobname.toc
\catcode`\@=12\bigbreak\bigskip}\fi}
\catcode`\@=12 % at signs are no longer letters
%
%	Unpleasantness in calling in abstract and title fonts
\edef\tfontsize{\ifx\answ\bigans scaled\magstep3\else scaled\magstep4\fi}
\font\titlerm=cmr10 \tfontsize \font\titlerms=cmr7 \tfontsize
\font\titlermss=cmr5 \tfontsize \font\titlei=cmmi10 \tfontsize
\font\titleis=cmmi7 \tfontsize \font\titleiss=cmmi5 \tfontsize
\font\titlesy=cmsy10 \tfontsize \font\titlesys=cmsy7 \tfontsize
\font\titlesyss=cmsy5 \tfontsize \font\titleit=cmti10 \tfontsize
\skewchar\titlei='177 \skewchar\titleis='177 \skewchar\titleiss='177
\skewchar\titlesy='60 \skewchar\titlesys='60 \skewchar\titlesyss='60
\def\titlefont{\def\rm{\fam0\titlerm}% switch to title font
\textfont0=\titlerm \scriptfont0=\titlerms \scriptscriptfont0=\titlermss
\textfont1=\titlei \scriptfont1=\titleis \scriptscriptfont1=\titleiss
\textfont2=\titlesy \scriptfont2=\titlesys \scriptscriptfont2=\titlesyss
\textfont\itfam=\titleit \def\it{\fam\itfam\titleit}\rm}
 \ifx\answ\bigans\else scaled\magstep1\fi
\ifx\answ\bigans\def\abstractfont{\tenpoint}\else
\font\absit=cmti10 scaled \magstep1
\font\abssl=cmsl10 scaled \magstep1
\font\absrm=cmr10 scaled\magstep1 \font\absrms=cmr7 scaled\magstep1
\font\absrmss=cmr5 scaled\magstep1 \font\absi=cmmi10 scaled\magstep1
\font\absis=cmmi7 scaled\magstep1 \font\absiss=cmmi5 scaled\magstep1
\font\abssy=cmsy10 scaled\magstep1 \font\abssys=cmsy7 scaled\magstep1
\font\abssyss=cmsy5 scaled\magstep1 \font\absbf=cmbx10 scaled\magstep1
\skewchar\absi='177 \skewchar\absis='177 \skewchar\absiss='177
\skewchar\abssy='60 \skewchar\abssys='60 \skewchar\abssyss='60
\def\abstractfont{\def\rm{\fam0\absrm}% switch to abstract font
\textfont0=\absrm \scriptfont0=\absrms \scriptscriptfont0=\absrmss
\textfont1=\absi \scriptfont1=\absis \scriptscriptfont1=\absiss
\textfont2=\abssy \scriptfont2=\abssys \scriptscriptfont2=\abssyss
\textfont\itfam=\absit \def\it{\fam\itfam\absit}\def\footnotefont{\tenpoint}%
\textfont\slfam=\abssl \def\sl{\fam\slfam\abssl}%
\textfont\bffam=\absbf \def\bf{\fam\bffam\absbf}\rm}\fi
\def\tenpoint{\def\rm{\fam0\tenrm}% switch back to 10-point type
\textfont0=\tenrm \scriptfont0=\sevenrm \scriptscriptfont0=\fiverm
\textfont1=\teni  \scriptfont1=\seveni  \scriptscriptfont1=\fivei
\textfont2=\tensy \scriptfont2=\sevensy \scriptscriptfont2=\fivesy
\textfont\itfam=\tenit \def\it{\fam\itfam\tenit}\def\footnotefont{\ninepoint}%
\textfont\bffam=\tenbf \def\bf{\fam\bffam\tenbf}\def\sl{\fam\slfam\tensl}\rm}
\font\ninerm=cmr9 \font\sixrm=cmr6 \font\ninei=cmmi9 \font\sixi=cmmi6
\font\ninesy=cmsy9 \font\sixsy=cmsy6 \font\ninebf=cmbx9
\font\nineit=cmti9 \font\ninesl=cmsl9 \skewchar\ninei='177
\skewchar\sixi='177 \skewchar\ninesy='60 \skewchar\sixsy='60
\def\ninepoint{\def\rm{\fam0\ninerm}% switch to footnote font
\textfont0=\ninerm \scriptfont0=\sixrm \scriptscriptfont0=\fiverm
\textfont1=\ninei \scriptfont1=\sixi \scriptscriptfont1=\fivei
\textfont2=\ninesy \scriptfont2=\sixsy \scriptscriptfont2=\fivesy
\textfont\itfam=\ninei \def\it{\fam\itfam\nineit}\def\sl{\fam\slfam\ninesl}%
\textfont\bffam=\ninebf \def\bf{\fam\bffam\ninebf}\rm}
%
%---------------------------------------------------------------------
%
\def\noblackbox{\overfullrule=0pt}
\hyphenation{anom-aly anom-alies coun-ter-term coun-ter-terms}
\def\inv{^{\raise.15ex\hbox{${\scriptscriptstyle -}$}\kern-.05em 1}}

\def\Dsl{\,\raise.15ex\hbox{/}\mkern-13.5mu D} %this one can be subscripted
\def\dsl{\raise.15ex\hbox{/}\kern-.57em\partial}

\def\tr{{\rm tr}} \def\Tr{{\rm Tr}}
 %pound sterling
\def\lspace{\ifx\answ\bigans{}\else\qquad\fi}
\def\lbspace{\ifx\answ\bigans{}\else\hskip-.2in\fi} % $$\lbspace...$$
\def\boxeqn#1{\vcenter{\vbox{\hrule\hbox{\vrule\kern3pt\vbox{\kern3pt
	\hbox{${\displaystyle #1}$}\kern3pt}\kern3pt\vrule}\hrule}}}
\def\mbox#1#2{\vcenter{\hrule \hbox{\vrule height#2in
		\kern#1in \vrule} \hrule}}  %e.g. \mbox{.1}{.1}
%	matters of taste
%\def\tilde{\widetilde} \def\bar{\overline} \def\hat{\widehat}
%
% some sample definitions
  %     curly letters

\def\vev#1{\langle #1 \rangle}

\def\darr#1{\raise1.5ex\hbox{$\leftrightarrow$}\mkern-16.5mu #1}
 %pound sterling

 %puts a small half in a displayed eqn
\def\roughly#1{\raise.3ex\hbox{$#1$\kern-.75em\lower1ex\hbox{$\sim$}}}

%%%%%%%%%%%%%%%%%%%%%%%%%%%%%%%%%%%%%%%%%%%%%%%%%%%%%%%%%%%%%%%%%%%%%
%%%%%%%%%%%%%%%   Subsubsection  %%%%%%%%%%%%%%%%%%%%%%%%%%%%%%%%%%%%
%%%%%%%%%%%%%%%%%%%%%%%%%%%%%%%%%%%%%%%%%%%%%%%%%%%%%%%%%%%%%%%%%%%%%
\global\newcount\subsubsecno \global\subsubsecno=0
\def\subsubsec#1{\global\advance\subsubsecno by1%
{\toks0{#1}\message{(\the\secno\the\subsecno\the\subsubsecno. \the\toks0)}}%
\ifnum\lastpenalty>9000\else\bigbreak\fi
\noindent{\it\hyperdef\hypernoname{subsubsection}{\the\secno.\the\subsecno\the\subsubsecno}%
{\the\secno.\the\subsecno.\the\subsubsecno.} #1}
%%% Add Subsubsections to Index
%% \writetoca{\string\quad{\string\hyperref{}{subsubsection}{\the\secno\the\subsecno\the
%%\subsubsecno}{\baselineskip=9pt\it\the\secno.\the\subsecno.\the\subsubsecno.}}
%% {\baselineskip=9pt\it\ #1}}
\par\nobreak\medskip\nobreak}
%%%%%%%%%%%%%%%%%%%%%%%%%%%%%%%%%%%%%%%%%%%%%%%%%%%%%%%%%%%%%%%%%%%%%
%%%%%%%%%%%%%%%%%%%%%%%%%%%%%%%%%%%%%%%%%%%%%%%%%%%%%%%%%%%%%%%%%%%
%%%%%% BOX
%%%%%%%%%%%%%%%%%%%%%%%%%%%%%%%%%%%%%%%%%%
\def\boxit#1{\vbox{\hrule\hbox{\vrule\kern8pt
\vbox{\hbox{\kern8pt}\hbox{\vbox{#1}}\hbox{\kern8pt}}
\kern8pt\vrule}\hrule}}
\def\mathboxit#1{\vbox{\hrule\hbox{\vrule\kern8pt\vbox{\kern8pt
\hbox{$\displaystyle #1$}\kern8pt}\kern8pt\vrule}\hrule}}
%%%%%%%%%%%%%%%%%%%%%%%%%%%%%%%%%%%%%%%%%%%%%%%%%%%%%%%%%%%%%%%%%%%
%%%%%%%%%%%%%%%%%%%%%%%%%%%%%%%%%%%%%%%%%%%%%%%%%%%%%%%%%%%%%%%%
%%%%%   Dirac-Slash
%%%%%%%%%%%%%%%%%%%%%%%%%%%%%%%%%%%%%%%%%%%%%%%%%%%%%%%%%%%%%%%%
\def\slashchar#1{\setbox0=\hbox{$#1$}           % set a box for #1
   \dimen0=\wd0                                 % and get its size
   \setbox1=\hbox{/} \dimen1=\wd1               % get size of /
   \ifdim\dimen0>\dimen1                        % #1 is bigger
      \rlap{\hbox to \dimen0{\hfil/\hfil}}      % so center / in box
      #1                                        % and print #1
   \else                                        % / is bigger
      \rlap{\hbox to \dimen1{\hfil$#1$\hfil}}   % so center #1
      /                                         % and print /
   \fi}
%%%%%%%%%%%%%%%%%%%%%%%%%%%%%%%%%%%%%%%%%%%%%%%%%%%%%%%%%%%%%%%%%
%%%%%%%%%%%%%%%%%%%%%%%%%%%%%%%%%%%%%%%%%%%%%%%%%%%%%%%%%%%
%  To produce a box for a Dalembertian (adapted from p. 320 of TeXbook):
\def\sqr#1#2{{\vcenter{\vbox{\hrule height.#2pt
         \hbox{\vrule width.#2pt height#1pt \kern#1pt
            \vrule width.#2pt}
         \hrule height.#2pt}}}}

%%%%%%%%%%%%%%%%%%%%%%%%%%%%%%%%%%%%%%%%%%%%%%%%%%%%%%%%%%%

%%%%%%%%%%%%%%%%%%%%%%%%%%%%%%%%%%%%%%%%
%%%%%%%%%%%  load AMS-fonts cf. TeX book p.158  ** can be downloaded from http://arxiv.org/macros/
\input amssym.def
\input amssym.tex
%%% e.g.: $\frak A$  $\Bbb  A$
%%%\input amstex.tex
%%%  e.g.: $\Cal A$
%%%\let\footnote\plainfootnote
%%%%%%%%%%%%%%%%%%%%%%%%%%%%%%%%%%%%%%%%%%%%%%%%%%%%%%%%%%%%%%%%%%%%%
\noblackbox
%\draftmode  %%%%%%%%%%%%%%%% Lineskip
%%%%%%%%%%%%%%%%%%%%%%%%%%%%%%%%%%%%%%%
\baselineskip=14.5pt
%%%%%%%%%%%% Local definitions %%%%%%%%%%%%%%%%%%%%%%%%%%%%%%%%%%%%
\def\crr{\noalign{\vskip5pt}}

\def\comment#1{{}}
\def\ss#1{{\scriptstyle{#1}}}

\def\z{{\zeta}}
\def\ap{\alpha'}

\def\cf{{\it cf.\ }}
\def\ie{{\it i.e.\ }}
\def\eg{{\it e.g.\ }}
\def\eqq{{\it Eq.\ }}
\def\eqqs{{\it Eqs.\ }}

\def\al{\alpha}

\def\si{\sigma}

\def\bet{\beta}

%%%%%%%%%%%%%%%%%%%%%%%%%%%%%%%%%%%%%%%%%%%%%%%%%%%%%%%%%%%%%%%%%
%%%%% Referencing  %%%%%%%%%%%%%%%%%%%%%%%%%%%%%%%%%%%%%%%%%%%%%%
%%%%%%%%%%%%%%%%%%%%%%%%%%%%%%%%%%%%%%%%%%%%%%%%%%%%%%%%%%%%%%%%%
\newif\ifnref

\nreffalse
%%%%%%%%%%%%%%%%%%%%%%%%%%%%%%%%%%%%%%%%%%%%%%%%%%%%%%%%%%%%%%%%%%
%%%%%%%%%%%%%%%%%   Stuff for Figures  %%%%%%%%%%%%%%%%%%%%%%%%%%%
%%%%%%%%%%%%%%%%%%%%%%%%%%%%%%%%%%%%%%%%%%%%%%%%%%%%%%%%%%%%%%%%%%

\input epsf

\def\figin{\epsfcheck\figin}\def\figins{\epsfcheck\figins}
\def\epsfcheck{\ifx\epsfbox\UnDeFiNeD
\message{(NO epsf.tex, FIGURES WILL BE IGNORED)}
\gdef\figin##1{\vskip2in}\gdef\figins##1{\hskip.5in}% blank space instead
\else\message{(FIGURES WILL BE INCLUDED)}%
\gdef\figin##1{##1}\gdef\figins##1{##1}\fi}
\def\DefWarn#1{}
\def\figinsert{\goodbreak\midinsert}  % instead \topinsert
\def\ifig#1#2#3{\DefWarn#1\xdef#1{Fig.~\the\figno}
\writedef{#1\leftbracket fig.\noexpand~\the\figno}%
\figinsert\figin{\centerline{#3}}\medskip\centerline{\vbox{\baselineskip12pt
\advance\hsize by -1truein\noindent\footnotefont\centerline{{\bf
Fig.~\the\figno}\ \sl #2}}}
\bigskip\endinsert\global\advance\figno by1}

%%%%%%%%  Second line in Figure caption
\def\iifig#1#2#3#4{\DefWarn#1\xdef#1{Fig.~\the\figno}
\writedef{#1\leftbracket fig.\noexpand~\the\figno}%
\figinsert\figin{\centerline{#4}}\medskip\centerline{\vbox{\baselineskip12pt
\advance\hsize by -1truein\noindent\footnotefont\centerline{{\bf
Fig.~\the\figno}\ \ \sl #2}}}\smallskip\centerline{\vbox{\baselineskip12pt
\advance\hsize by -1truein\noindent\footnotefont\centerline{\ \ \ \sl #3}}}
\bigskip\endinsert\global\advance\figno by1}

%%%%%%%%%%%%%%%%%%%%%%%%%%%%%%%%%%%%%%%%%%%%%%%%%%%%%%%%%%%%%%%%%%%%%
%%%%%%%%%%%%%%%   Standard alltime definitions   %%%%%%%%%%%%%%%%%%%%
%%%%%%%%%%%%%%%%%%%%%%%%%%%%%%%%%%%%%%%%%%%%%%%%%%%%%%%%%%%%%%%%%%%%%

\def\tilde{\widetilde}

\def\h {{1\over 2}}

\def\ov {\overline}
\def\o {\over}
\def\fc#1#2{{#1 \o #2}}

\def\IZ{ {\bf Z}}\def\IP{{\bf P}}\def\IC{{\bf C}}\def\IR{ {\bf R}}
\def\IN{ {\bf N}}
\def\IQ{ {\bf Q}}

      % For Eisenstein E2
  % For Polylogarithm

\def\br{\hfill\break}
\def\tr {{\rm tr}}

\def\lf {\left}
\def\ri {\right}
\def\ra {\rightarrow}

\def\p {\partial}

 \def\Sc {{\cal S}}
\def\Mc {{\cal M}} \def\Ac {{\cal A}}
 
 \def\Uc {{\cal U}}
 
 \def\Nc {{\cal N}}

\def\Hc{{\cal H}}

\def\Af{{\frak A}}
%%\def\Acb{{\pmb{\Cal A}}}

%%%%%%%%  My Shuffle Product
\def\shuffle{{\hskip0.10cm \vrule height 0pt width 8pt depth 0.75pt  \hskip-0.3cm\ss{\rm III}\hskip0.05cm}}
%%%%%%%%%%%%%%%%%%%%%%%%%%%%%%%%%%%%%%%%%%%%%%%%%%%%%%%%%%%%%%%%%%%
\def\sv{{\rm sv}}
\def\SVM{{\zeta^m_{\rm sv}}}
\def\SV{{\zeta_{\rm sv}}}

%%%%%%%%%%%%%%%%%%%%%%%%%%%%%%%%%%%%%%%%%%%%%%%%%%%%%%%%%%%%%%%%

\lref\StiebergerWEA{
  S.~Stieberger,
``Closed Superstring Amplitudes, Single-Valued Multiple Zeta Values and Deligne Associator,''
[arXiv:1310.3259 [hep-th]].
%%CITATION = MPP-2013-278%%
}

\lref\Deligne{
P. Deligne,
``Le groupe fondamental de la droite projective moins trois points,''
in: Galois groups over $\IQ$, Springer, MSRI publications {\bf 16} (1989), 72-297;
``Periods for the fundamental group,''
Arizona Winter School 2002.}

\lref\Goncharov{
A.B. Goncharov,
``Galois symmetries of fundamental groupoids and noncommutative geometry,''
 Duke Math. J. 128 (2005) 209-284. [arXiv:math/0208144v4 [math.AG]];\br
F.~Brown, ``Mixed Tate Motives over $\IZ$,'' Ann. Math. 175 (2012) 949--976.
}

\lref\Broedel{
J.~Broedel, O.~Schlotterer and S.~Stieberger,
 ``Polylogarithms, Multiple Zeta Values and Superstring Amplitudes,''
Fortsch.\  Phys.\  {\bf 61}, 812 (2013).
[arXiv:1304.7267 [hep-th]].
%%CITATION = DAMTP-2013-22%%
}

\lref\SS{
  O.~Schlotterer and S.~Stieberger,
``Motivic Multiple Zeta Values and Superstring Amplitudes,''
J.\ Phys.\ A {\bf 46}, 475401 (2013).
[arXiv:1205.1516 [hep-th]].
%%CITATION = arXiv:1205.1516%%
}

\lref\SVMZV{
  F. Brown,
``Single-valued periods and multiple zeta values,''
[arXiv:1309.5309 [math.NT]].
%%CITATION = arXiv:1309.5309%%
}

\lref\StiebergerFH{
  S.~Stieberger and T.R.~Taylor,
  ``Non-Abelian Born-Infeld action and type I - heterotic duality I: Heterotic $F^6$ terms at two loops,''
Nucl.\ Phys.\ B {\bf 647}, 49 (2002).
[hep-th/0207026];
%%CITATION = hep-th/0207026%%
``Non-Abelian Born-Infeld action and type I - heterotic duality II: Nonrenormalization theorems,''
Nucl.\ Phys.\ B {\bf 648}, 3 (2003).
[hep-th/0209064].
%%CITATION = hep-th/0209064%%
}

\lref\TseytlinFY{
  A.A.~Tseytlin,
  ``On SO(32) heterotic type I superstring duality in ten-dimensions,''
Phys.\ Lett.\ B {\bf 367}, 84 (1996).
[hep-th/9510173]; ``Heterotic type I superstring duality and low-energy effective actions,''
Nucl.\ Phys.\ B {\bf 467}, 383 (1996).
[hep-th/9512081].
%%CITATION = hep-th/9510173%%
}

\lref\HullYS{
  C.M.~Hull and P.K.~Townsend,
  ``Unity of superstring dualities,''
Nucl.\ Phys.\ B {\bf 438}, 109 (1995).
[hep-th/9410167].
%%CITATION = hep-th/9410167%%
}

\lref\BrownPoly{
F. Brown,
``Single-valued multiple polylogarithms in one variable,''
C.R. Acad. Sci. Paris, Ser. I {\bf 338}, 527-532 (2004).}

\lref\Schnetz{
  O.~Schnetz,
``Graphical functions and single-valued multiple polylogarithms,''\br
[arXiv:1302.6445 [math.NT]].
%%CITATION = arXiv:1302.6445%%
}

\lref\StiebergerHZA{
  S.~Stieberger and T.R.~Taylor,
``Superstring Amplitudes as a Mellin Transform of Supergravity,''
Nucl.\ Phys.\ B {\bf 873}, 65 (2013)
[arXiv:1303.1532 [hep-th]];
%%CITATION = MPP--2013--18%%
``Superstring/Supergravity Mellin Correspondence in Grassmannian Formulation,''
Phys.\ Lett.\ B {\bf 725}, 180 (2013)
[arXiv:1306.1844 [hep-th]].
%%CITATION = MPP-2013-148%%
}

\lref\WittenEX{
  E.~Witten,
  ``String theory dynamics in various dimensions,''
Nucl.\ Phys.\ B {\bf 443}, 85 (1995).
[hep-th/9503124].
%%CITATION = hep-th/9503124%%
}

\lref\PolchinskiDF{
  J.~Polchinski and E.~Witten,
``Evidence for heterotic - type I string duality,''
Nucl.\ Phys.\ B {\bf 460}, 525 (1996).
[hep-th/9510169].
%%CITATION = hep-th/9510169%%
}

\lref\StiebergerHQ{
  S.~Stieberger,
``Open \& Closed vs. Pure Open String Disk Amplitudes,''
[arXiv:0907.2211 [hep-th]].
%%CITATION = arXiv:0907.2211%%
}

\lref\BjerrumBohrRD{
  N.E.J.~Bjerrum-Bohr, P.H.~Damgaard and P.~Vanhove,
  ``Minimal Basis for Gauge Theory Amplitudes,''
  Phys.\ Rev.\ Lett.\  {\bf 103}, 161602 (2009)
  [arXiv:0907.1425 [hep-th]].
  %%CITATION = PRLTA,103,161602;%%
}

\lref\Brown{
  F.~Brown,
``On the decomposition of motivic multiple zeta values,''
 in `Galois-Teichm\"uller Theory and Arithmetic Geometry', Advanced Studies in Pure Mathematics 63 (2012) 31-58 [arXiv:1102.1310 [math.NT]].
%%CITATION = arXiv:1102.1310%%
}

\lref\GRAV{
  S.~Stieberger,
 ``Constraints on Tree-Level Higher Order Gravitational Couplings in Superstring Theory,''
Phys.\ Rev.\ Lett.\  {\bf 106}, 111601 (2011)
[arXiv:0910.0180 [hep-th]].
%%CITATION = arXiv:0910.0180%%
}

\lref\MafraKJ{
  C.R.~Mafra, O.~Schlotterer and S.~Stieberger,
``Explicit BCJ Numerators from Pure Spinors,''
JHEP {\bf 1107}, 092 (2011).
[arXiv:1104.5224 [hep-th]].
%%CITATION = arXiv:1104.5224%%
}

\lref\Bohr{
  N.E.J.~Bjerrum-Bohr, P.H.~Damgaard, T.~Sondergaard and P.~Vanhove,
 ``The Momentum Kernel of Gauge and Gravity Theories,''
JHEP {\bf 1101}, 001 (2011).
[arXiv:1010.3933 [hep-th]].
%%CITATION = arXiv:1010.3933%%
}

\lref\BernSV{
  Z.~Bern, L.J.~Dixon, M.~Perelstein and J.S.~Rozowsky,
``Multileg one loop gravity amplitudes from gauge theory,''
Nucl.\ Phys.\ B {\bf 546}, 423 (1999).
[hep-th/9811140].
%%CITATION = hep-th/9811140%%
}

\lref\KawaiXQ{
  H.~Kawai, D.C.~Lewellen and S.H.H.~Tye,
``A Relation Between Tree Amplitudes Of Closed And Open Strings,''
  Nucl.\ Phys.\  B {\bf 269}, 1 (1986).
  %%CITATION = NUPHA,B269,1;%%
}

\lref\BCJ{
Z.~Bern, J.J.M.~Carrasco and H.~Johansson,
  ``New Relations for Gauge-Theory Amplitudes,''
  Phys.\ Rev.\  D {\bf 78}, 085011 (2008)
  [arXiv:0805.3993 [hep-ph]].
  %%CITATION = PHRVA,D78,085011;%%
}

\lref\BCJi{
Z.~Bern, J.J.M.~Carrasco and H.~Johansson,
``Perturbative Quantum Gravity as a Double Copy of Gauge Theory,''
Phys.\ Rev.\ Lett.\  {\bf 105}, 061602 (2010)
[arXiv:1004.0476 [hep-th]];\br
%%CITATION = arXiv:1004.0476%%
Z.~Bern, T.~Dennen, Y.-t.~Huang and M.~Kiermaier,
``Gravity as the Square of Gauge Theory,''
Phys.\ Rev.\ D {\bf 82}, 065003 (2010)
[arXiv:1004.0693 [hep-th]].
%%CITATION = arXiv:1004.0693%%
}

\lref\CachazoIEA{
  F.~Cachazo, S.~He and E.Y.~Yuan,
``Scattering of Massless Particles: Scalars, Gluons and Gravitons,''
[arXiv:1309.0885 [hep-th]].
%%CITATION = arXiv:1309.0885%%
}

\lref\Duhr{
  S.~Leurent and D.~Volin,
``Multiple zeta functions and double wrapping in planar N=4 SYM,''
Nucl.\ Phys.\ B {\bf 875}, 757 (2013).
[arXiv:1302.1135 [hep-th]];\br
%%CITATION = NORDITA-2013-11%%
L.J.~Dixon, C.~Duhr and J.~Pennington,
 ``Single-valued harmonic polylogarithms and the multi-Regge limit,''
JHEP {\bf 1210}, 074 (2012).
[arXiv:1207.0186 [hep-th]];\br
%%CITATION = arXiv:1207.0186%%
F.~Chavez and C.~Duhr,
``Three-mass triangle integrals and single-valued polylogarithms,''
JHEP {\bf 1211}, 114 (2012).
[arXiv:1209.2722 [hep-ph]];\br
%%CITATION = arXiv:1209.2722%%
V.~Del Duca, L.J.~Dixon, C.~Duhr and J.~Pennington,
``The BFKL equation, Mueller-Navelet jets and single-valued harmonic polylogarithms,''
[arXiv:1309.6647 [hep-ph]].
%%CITATION = arXiv:1309.6647%%
}

\lref\DolanEH{I.B. Frenkel and Y. Zhu,
``Vertex Operator Algebras Associated to Representations of Affine and Virasoro
Algebras,''
Duke Math J. {\bf 66}, 123 (1992);\br
  L.~Dolan and P.~Goddard,
``Current Algebra on the Torus,''
Commun.\ Math.\ Phys.\  {\bf 285}, 219 (2009).
[arXiv:0710.3743 [hep-th]].
%%CITATION = arXiv:0710.3743%%
}

\lref\ElvangCUA{
  H.~Elvang and Y.-t.~Huang,
``Scattering Amplitudes,''
[arXiv:1308.1697 [hep-th]].
%%CITATION = arXiv:1308.1697%%
}

\lref\MafraNVi{
  C.R.~Mafra, O.~Schlotterer and S.~Stieberger,
``Complete N-Point Superstring Disk Amplitude I. Pure Spinor Computation,''
Nucl.\ Phys.\ B {\bf 873}, 419 (2013).
[arXiv:1106.2645 [hep-th]].
%%CITATION = arXiv:1106.2645%%
}

\lref\MafraNVii{
  C.R.~Mafra, O.~Schlotterer and S.~Stieberger,
``Complete N-Point Superstring Disk Amplitude II. Amplitude and Hypergeometric Function Structure,''
Nucl.\ Phys.\ B {\bf 873}, 461 (2013).
[arXiv:1106.2646 [hep-th]].
%%CITATION = arXiv:1106.2646%%
}

\lref\LSW{
  W.~Lerche, A.N.~Schellekens and N.P.~Warner,
``Lattices and Strings,''
Phys.\ Rept.\  {\bf 177}, 1 (1989).
%%CITATION = CERN-TH-5155-88%%
}

\lref\BershadskyTA{
  M.~Bershadsky, S.~Cecotti, H.~Ooguri and C.~Vafa,
``Holomorphic anomalies in topological field theories,''
Nucl.\ Phys.\ B {\bf 405}, 279 (1993).
[hep-th/9302103].
%%CITATION = hep-th/9302103%%
}

\lref\KaplunovskyJW{
  V.~Kaplunovsky and J.~Louis,
``On Gauge couplings in string theory,''
Nucl.\ Phys.\ B {\bf 444}, 191 (1995).
[hep-th/9502077].
%%CITATION = hep-th/9502077%%
}

\lref\TyeDD{
  S.H.H. Tye and Y.~Zhang,
``Dual Identities inside the Gluon and the Graviton Scattering Amplitudes,''
JHEP {\bf 1006}, 071 (2010), [Erratum-ibid.\  {\bf 1104}, 114 (2011)].
[arXiv:1003.1732 [hep-th]];\br
%%CITATION = arXiv:1003.1732%%
N.E.J.~Bjerrum-Bohr, P.H.~Damgaard, T.~Sondergaard and P.~Vanhove,
``Monodromy and Jacobi-like Relations for Color-Ordered Amplitudes,''
JHEP {\bf 1006}, 003 (2010).
[arXiv:1003.2403 [hep-th]].
%%CITATION = arXiv:1003.2403%%
}

\lref\OchirovXBA{
  A.~Ochirov and P.~Tourkine,
``BCJ duality and double copy in the closed string sector,''
[arXiv:1312.1326 [hep-th]].
%%CITATION = IPHT-T13-196%%
}

\lref\GomezWZA{
  H.~Gomez and E.Y.~Yuan,
``N-Point Tree-Level Scattering Amplitude in the New Berkovits' String,''
[arXiv:1312.5485 [hep-th]].
%%CITATION = arXiv:1312.5485%%
}

\lref\BerkovitsXBA{
  N.~Berkovits,
``Infinite Tension Limit of the Pure Spinor Superstring,''
[arXiv:1311.4156 [hep-th]].
%%CITATION = ICTP-SAIFR-2013-13%%
}

\lref\GrossRR{
  D.J.~Gross, J.A.~Harvey, E.J.~Martinec and R.~Rohm,
``Heterotic String Theory. 2. The Interacting Heterotic String,''
Nucl.\ Phys.\ B {\bf 267}, 75 (1986).
%%CITATION = Print-85-0694 (PRINCETON)%%
}

%%%%%%%%%%%%%%%%%%%%%%%%%%%%%%%%%%%%%%%%%%%%%%%%%%%%%%%%%%%%%%%%%%%
\Title{\vbox{\rightline{MPP--2014--001}
}}
{\vbox{\centerline{Closed String Amplitudes as }\br
\centerline{Single--Valued Open String  Amplitudes}}}
\medskip
\centerline{Stephan Stieberger$^a$ and Tomasz R. Taylor$^b$}
\bigskip
\centerline{\it $^a$ Max--Planck--Institut f\"ur Physik}
\centerline{\it Werner--Heisenberg--Institut, 80805 M\"unchen, Germany}
\medskip
\centerline{\it  $^b$ Department of Physics}
\centerline{\it  Northeastern University, Boston, MA 02115, USA}

\vskip15pt

\medskip
\bigskip\bigskip\bigskip
\centerline{\bf Abstract}
\vskip .2in
\noindent

\noindent
We show that the single trace heterotic $N$--point
 tree--level gauge amplitude $\Ac^{\rm HET}_N$ can be obtained from the corresponding type I
amplitude $\Ac^{\rm I}_N$ by the single--valued ($\sv$) projection:
$\Ac^{\rm HET}_N=\sv(\Ac^{\rm I}_N)$.
This projection maps multiple zeta values to single--valued multiple zeta values.
The latter represent a subclass of multiple zeta values originating from single--valued multiple polylogarithms at unity.
Similar relations between open and closed string amplitudes or amplitudes of different string vacua
can be established. As a consequence the $\ap$--expansion of a closed string amplitude is dictated by that of the corresponding open string amplitude.
The combination of single--valued projections, Kawai--Lewellen--Tye relations and Mellin correspondence reveal a unity of all tree--level open and closed superstring amplitudes together with the maximally supersymmetric Yang--Mills and supergravity theories.
%Our relations give rise to give rise to a deeper connection between open and closed string amplitudes %than what is implied by Kawai--Lewellen--Tye relations.

\Date{}
\noindent
\goodbreak
%\listtoc
%\writetoc
\break
%%%%%%%%%%%%%%%%%%%%%%%%%%%%%%%%%%%%%%%%%%%%%%%%%%%%%%%%%%%%%%%%%%%%%%%%%%%%%%%
\newsec{Introduction}
\def\per{{\rm per}}

Perturbative open and closed string amplitudes seem to be rather different due to their underlying different world--sheet topologies. On the other hand,
mathematical methods entering their computations reveal some unexpected connections.
{}For example, at the string tree--level, the Kawai--Lewellen--Tye (KLT) relations
expose a non--trivial connection between open  and closed  superstring amplitudes \KawaiXQ.
Similarly, tree--level scattering involving both open  and closed strings can  entirely be described by
open string amplitudes only \StiebergerHQ.
Furthermore, recently in {\it Ref.} \StiebergerWEA\ a new relation between open (type I) and closed (type~II) superstring amplitudes has been found. The $\ap$--expansion (with $\ap$ being the inverse string tension) of the closed superstring amplitude can be cast into the same algebraic form as the open superstring amplitude: the closed superstring amplitude  is essentially the single--valued (\sv) version of the open superstring amplitude.

All these encountered string properties of scattering amplitudes suggest that there is a
deeper connection between perturbative gauge and gravity theories than expected.
Furthermore, these observations in string theory are attended by results in field theory. There an apparent similarity  between perturbative gauge-- and gravity--theories
is established  through the double copy construction \refs{\BCJ,\BCJi}. The latter can indeed be furnished by considering gauge amplitudes in heterotic string theory  as a simple consequence of their
underlying world--sheet structure \refs{\TyeDD,\OchirovXBA}.
Besides, recently in {\it Ref.} \CachazoIEA\ an interesting uniform description of field--theoretical gauge-- and gravity amplitudes has been presented  and we shall see that  some of its underlying structure is carried by the gauge amplitudes in heterotic string theory.

Yang Mills (YMs) and supergravity amplitudes appear in  the $\alpha' \ra 0$ limit of superstring theory. The recently discovered Mellin correspondence \StiebergerHZA\ allows (re)constructing the full--fledged tree--level open superstring amplitudes (to all orders in $\alpha'$) from this field--theory limit. In this work, we  employ the single--valued projection as a link between type~I and heterotic string theory. Together with the sv connection between type I and type II, Mellin correspondence and KLT relations, it appears in the web of connections between full--fledged string and field--theoretical tree--level amplitudes. The basic building blocks of superstring amplitudes are supplied by gauge theory describing the string zero modes.

The paper is organized as follows.
In Section 2 we review  the subspace of single--valued multiple zeta values, in which a large class of
  complex sphere integrals of closed superstring amplitudes lives. After exhibiting results on the tree--level open superstring $N$--point disk amplitude in Section 3 we compute the analogous heterotic
single trace $N$--gauge boson amplitude in a generic heterotic string vacuum.
We shall find, that the two amplitudes are related by the sv projection.
As a byproduct we verify that  heterotic string subamplitudes fulfil the same
amplitude relations, \ie Kleiss--Kuijf (KK)  and Bern--Carrasco--Johansson (BCJ)  relations as field--theory subamplitudes. In Section 4 we discuss   closed superstring amplitudes
involving gravitons or scalars and relate them to open superstring amplitudes.
In Section~5 we elucidate the  generic basic building blocks of
open and closed string world--sheet integrals and establish a relation between them. This correlation  
is responsible for the connection between open and closed superstring amplitudes. In Section 6, we discuss connections between string and field--theoretical amplitudes and comment on their relation to the web of string dualities.

\newsec{Single--valued multiple zeta--values  in superstring theory}

The analytic dependence on the inverse string tension~$\ap$ of string tree--level amplitudes
furnishes an extensive and rich mathematical structure, which is related to modern developments in number theory and arithmetic algebraic geometry, \cf  {\it Refs.} \refs{\SS,\StiebergerWEA}.

The topology of the string world--sheet describing tree--level scattering of open strings is
a disk, while tree--level scattering of closed strings is  characterized by a complex sphere.
Open string amplitudes are expressed by integrals along the boundary of the world--sheet disk (real projective line) as
iterated (real) integrals on  $\IR\IP^1\backslash\{0,1,\infty\}$, whose values
(more precisely the coefficients in their power series expansion in $\ap$) are given by multiple zeta values (MZVs)
\eqn\MZV{
\zeta_{n_1,\ldots,n_r}:=\zeta(n_1,\ldots,n_r)=
\sum\limits_{0<k_1<\ldots<k_r}\ \prod\limits_{l=1}^r k_l^{-n_l}\ \ \ ,\ \ \ n_l\in\IN^+\ ,\ n_r\geq2\ ,}
with $r$ specifying the depth and $w=\sum_{l=1}^rn_l$ denoting
the weight of the MZV $\zeta_{n_1,\ldots,n_r}$.
On the other hand, closed string amplitudes  are given by integrals over the complex world--sheet sphere as iterated integrals on
$\IP^1\backslash\{0,1,\infty\}$ integrated independently on all choices of paths.
While in the $\ap$--expansion of open superstring tree--level amplitudes generically the whole space of MZVs \MZV\ enters  \refs{\GRAV,\SS}, closed superstring tree--level amplitudes exhibit only a subset of MZVs appearing in their $\ap$--expansion \refs{\GRAV,\SS}. This subclass can be identified \StiebergerWEA\ as single--valued multiple zeta values (SVMZVs)
\eqn\trivial{
\SV(n_1,\ldots,n_r)\in\IR}
originating from single--valued multiple polylogarithms (SVMPs) at unity \BrownPoly.
SVMZVs have  recently  been studied by Brown in \SVMZV\ from a mathematical point of view. They have been identified as the coefficients in an infinite series expansion of the Deligne associator \Deligne\
in two non--commutative variables. On the other hand, from a physical point of view SVMZVs
appear in the computation of graphical functions for certain Feynman amplitudes \Schnetz.

The numbers \trivial\ can be obtained from the MZVs \MZV\ by introducing the following homomorphism:
\eqn\mapSV{
\sv: \z_{n_1,\ldots,n_r}\mapsto\ \SV( n_1,\ldots,n_r )\ .}
The numbers \trivial\ satisfy the same double shuffle and associator relations than the
usual MZVs \MZV\ and many more relations \SVMZV.
For instance we have (\cf {\it Ref.} \StiebergerWEA\ for more examples):
\eqn\Example{\eqalign{
\sv(\z_2)&=\SV(2)=0\ ,\cr
\sv(\z_{2n+1})&=\SV(2n+1)=2\ \zeta_{2n+1}\ ,\ \ \ n\geq 1\ ,\cr
\sv(\z_{3,5})&=-10\ \z_3\ \z_5\ \ \ ,\ \ \ \sv(\z_{3,7})=-28\ \z_3\ \z_7-12\ \z_5^2\ ,\cr
\sv(\z_{3,3,5})&=2\ \z_{3,3,5}-5\ \z_3^2\ \z_5+90\ \z_2\ \z_9+\fc{12}{5}\ \z_2^2\ \z_7-\fc{8}{7}\ \z_2^3\ \z_5^2\ ,\ldots\ .}}

Strictly speaking, the map $\sv$ is defined in the Hopf algebra $\Hc$ of motivic MZVs $\zeta^m$.
Motivic MZVs\foot{Motivic aspects of MZVs have recently matured in describing tree--level open superstring amplitudes in {\it Ref.} \SS\ and tree--level closed superstring amplitudes in
{\it Ref.} \StiebergerWEA.}
 $\zeta^m$ are defined as elements of the algebra $\Hc=\bigoplus_{w\geq0}\Hc_w$ over $\IQ$, which is graded for the weight $w$ and equipped with
the  period homomorphism $\per: \Hc\ra \IR$, which maps $\zeta^m_{n_1,\ldots,n_r}$
to $\zeta_{n_1,\ldots,n_r}$, \ie $\per(\z^m_{n_1,\ldots,n_r})=\z_{n_1,\ldots,n_r}$ \Goncharov.
In the  algebra $\Hc$ the homomorphism $\sv: \Hc\ra\Hc^\sv$, with
\eqn\motivicSV{
\z^m_{n_1,\ldots,n_r}\mapsto\SVM(n_1,\ldots,n_r)\ ,}
and
\eqn\Withh{
\SVM(2)=0}
can be constructed \SVMZV.
The motivic SVMZVs $\SVM(n_1,\ldots,n_r)$ generate the subalgebra $\Hc^\sv$
of the Hopf algebra $\Hc$ and satisfy all motivic relations between MZVs.

In practice, the map $\sv$ is constructed recursively
in the (trivial) algebra--comodule  $\Uc=\Uc'\otimes_\IQ \IQ[f_2]$  with the
first factor $\Uc'=\IQ\vev{f_3,f_5,\ldots}$ generated by all non--commutative words in the letters $f_{2i+1}$ \Brown. We have $\Hc\simeq\Uc$, in particular $\z_{2i+1}\simeq f_{2i+1}$.  The homomorphism
\eqn\mapsv{
\sv:\Uc'\longrightarrow\Uc'\ ,}
with
\eqn\Mapsv{
w\longmapsto \sum_{uv=w} u\shuffle \tilde v\ ,}
and
\eqn\Withhh{
\sv(f_2)=0}
maps (\eg $\sv(f_{2i+1})=2f_{2i+1}$) the algebra of non--commutative words $w\in\Uc$ to the smaller subalgebra $\Uc^\sv$, which describes  the space of SVMZVs \SVMZV.
In \eqq \mapsv\ the word $\tilde v$ is the reversal of the word $v$ and $\shuffle$ is the shuffle product.
For more details we refer the reader to the original reference \SVMZV\ and subsequent
applications in~\StiebergerWEA.

In supersymmetric Yang--Mills (SYM) theory a large class of Feynman integrals in four space--time dimensions lives in the subspace of SVMZVs or SVMPs, \cf {\it Refs.} \Duhr.
As pointed out by Brown in \SVMZV, this fact opens the interesting possibility
to replace general amplitudes  with their single--valued versions (defined by the map $\sv$),
which should lead to considerable simplifications.
In string theory this simplification occurs by replacing open superstring amplitudes by their
single--valued versions describing closed superstring amplitudes.

%%%%
%%%%

\newsec{Heterotic gauge amplitudes as single--valued type I gauge amplitudes}

In this Section we shall compute single trace $N$--gluon tree--level subamplitudes in
heterotic string vacua. Our results will then be compared with the corresponding type I amplitudes.

Let us first review the latter. The open superstring $N$--gluon tree--level amplitude $\Af^{\rm I}_N$
decomposes into a sum
\eqn\DECO{
\Af^{\rm I}_N=(g_{YM}^{\rm I})^{N-2}\ \sum_{\Pi\in S_{N}/{\bf Z}_2}\Tr(T^{a_{\Pi(1)}}\ldots T^{a_{\Pi(N)}})\
\Ac^{\rm I}(\Pi(1),\ldots,\Pi(N))}
over color ordered subamplitudes $\Ac^{\rm I}(\Pi(1),\ldots,\Pi(N))$ supplemented by a group trace
in the fundamental representation. Above, the YM coupling is denoted by $g_{YM}^{\rm I}$, which
in type I superstring theory is given by $g_{YM}^{\rm I}\sim e^{\Phi/2}$ with the dilaton field $\Phi$.
The sum runs over all permutations $S_N$ of labels $i=1,\ldots,N$
modulo cyclic permutations ${\bf Z}_2$, which preserve the group trace.
A vector $\Ac^{\rm I}$ with its entries
$\Ac^{\rm I}_\pi=\Ac^{\rm I}(\pi)$  describing the
$(N-3)!$ independent open $N$--point superstring subamplitudes
\eqn\defA{
\Ac^{\rm I}(\pi):=\Ac^{\rm I}(1,\pi(2,\ldots,N-2),N-1,N)\ \ \ ,\ \ \ \pi\in S_{N-3}}
can be given by the  matrix expression $\Ac^I=FA$ \refs{\MafraNVi,\MafraNVii}. In components the latter reads
\eqn\OPEN{
\Ac^{\rm I}(\pi)=\sum_{\sigma\in S_{N-3}}
F_{\pi\sigma}\ A(\sigma)\ \ \ ,\ \ \ \pi\in S_{N-3}\ ,}
with  the $(N-3)!$ (independent) SYM subamplitudes
\eqn\subYM{
A(\si):=A(1,\si(2,\ldots,N-2),N-1,N)\ \ \ ,\ \ \ \si\in S_{N-3}\ ,}
which constitute the vector $A$ with entries $A_{\si}=A(\si)$ and the period matrix
$F_{\pi\sigma}$. According to \Broedel\ the latter can be expressed in terms of basic open string world--sheet disk integrals  ($z_{ij}=z_i-z_j$)
\eqnn\disk{
$$\eqalignno{
Z_\pi(\rho)&:= Z_\pi(1,\rho(2,\ldots,N-2),N,N-1)\cr
&=\lf(\prod_{j=2}^{N-2}
\int\limits_{D(\pi)} dz_j\ri) \   \fc{\prod\limits_{i<j}^{N-1} |z_{ij}|^{s_{ij}}}{  z_{1\rho(2)} z_{\rho(2),\rho(3)} \ldots z_{\rho(N-3),\rho(N-2)}} &\disk}$$}
as:
\eqn\Period{
F_{\pi\sigma}=(-1)^{N-3}\ \sum_{\rho\in S_{N-3}}Z_\pi(\rho)\ S[\rho|\sigma]\ .}
Above we have the KLT kernel\foot{The matrix $S$ with entries $S_{\rho,\si}=S[\rho|\si]$ is defined as a $(N-3)!\times (N-3)!$ matrix with its rows and columns corresponding to the orderings $\rho \equiv \{\rho(2),\ldots,\rho(N-2)\}$ and
$\si \equiv \{\si(2),\ldots,\si(N-2)\}$, respectively. The matrix $S$ is symmetric, \ie $S^t=S$.} \refs{\KawaiXQ,\BernSV,\Bohr}
\eqn\kernel{
S[\rho|\si]:=S[\, \rho(2,\ldots,N-2) \, | \, \si(2,\ldots,N-2) \, ] = \prod_{j=2}^{N-2} \Big( \, s_{1,j_\rho} \ + \ \sum_{k=2}^{j-1} \theta(j_\rho,k_\rho) \, s_{j_\rho,k_\rho} \, \Big)\ ,}
with $j_\rho=\rho(j)$ and  $\theta(j_\rho,k_\rho)=1$
if the ordering of the legs $j_\rho,k_\rho$ is the same in both orderings
$\rho(2,\ldots,N-2)$ and $\si(2,\ldots,N-2)$, and zero otherwise.
Due to conformal invariance on the world--sheet in \disk\ we have fixed the world--sheet positions  as $z_1=0,\ z_{N-1}=1,\ z_N=\infty$ and
the remaining $N-3$ points $z_i$ are integrated along the boundary of the disk
subject to the ordering
$D(\pi)=\{z_j\in \IR\ |\ 0<z_{\rho(2)}<\ldots<z_{\rho(N-2)}<1\}$.
Furthermore, we have the real numbers:
\eqn\Mandel{
s_{ij}=\ap (k_i+k_j)^2=2\ap k_ik_j\ .}
The latter describe the $\h N(N-3)$ independent kinematic invariants of the scattering process
involving $N$ external momenta $k_i,\ i=1,\ldots,N$ and $\ap$ is the inverse string tension. Note, that in the field--theory limit $\ap\ra0$ we have
\eqn\wehave{
\lf.Z\ri|_{\ap=0}=(-1)^{N-3}\ S^{-1}\ ,\ \ie \lf.F\ri|_{\ap=0}=1\ .}

Let us now move on to the gluon scattering in heterotic string vacua.
The string world--sheet describing the tree--level scattering of $N$
closed strings has the topology of a complex sphere with $N$ insertions
of  vertex operators. The string $S$--matrix elements can be computed from the expression
\eqn\TODO{
\Af_N^{\rm HET}=V_{\rm CKG}^{-1}\  \lf(\prod_{j=1}^{N}\int\limits_{z_j\in\IC} d^2z_j\ri) \
\vev{V(z_1,\ov z_1,\xi_1,k_1)\ldots V(z_N,\ov z_N,\xi_N,k_N)}}
involving $N$ gluon vertex operators  $V(z_l,\ov z_l,\xi_l,k_l)$.
In heterotic string vacua  in the RNS formalism the latter are given as \GrossRR
\eqnn\vertex{
$$\eqalignno{V^{(-1)}(z,\ov z,\xi,k)&=g_c\ \xi_\mu\ J^a( z)\ e^{-\phi(\ov z)}\ \psi^\mu(\ov z)\ e^{ik_\rho X^\rho(z,\ov  z)}\ ,&\vertex\cr
V^{(0)}( z,\ov z,\xi,k)&=g_c\lf(\fc{2}{\ap}\ri)^{1/2}\ \xi_\mu\ J^a( z)\ \lf[\ i\p X^\mu(\ov z)+\fc{\ap}{2}\ k_\lambda\psi^\lambda(\ov z)\ \psi(\ov  z)^\mu\ \ri]\ e^{ik_\rho X^\rho( z,\ov  z)}\ ,}$$}
in the $(-1)$-- and zero--ghost picture, respectively. The polarization $\xi_\mu$ satisfies the on--shell condition $\xi_\mu k^\mu=0$, with the external space--time momenta $k$ subject to the on--shell constraint $k_\mu k^\mu=0$. The heterotic closed string
coupling $g_c$ is given by $g_c=(4\pi)^{-1}\ap^{1/2} g_{YM}^{\rm HET}$, with the heterotic YM coupling  $g_{YM}^{\rm HET}\sim e^{\Phi}$.
In order to cancel the background ghost charge on the sphere, two vertices in the correlator
\TODO\ will be inserted in the $(-1)$--ghost picture, with the remaining ones in the
zero--ghost picture. Furthermore, in \TODO, the factor $V_{\rm CKG}$ accounts for the volume of
the conformal Killing group of the sphere after choosing the conformal gauge. It will be
canceled by fixing three vertex positions and introducing the respective $c$--ghost correlator.

Since we are discussing tree--level string amplitudes and shall not be concerned with world--sheet modular invariance or finiteness of string loop amplitudes, our findings hold for any space--time dimensions, any amount of supersymmetries ($\Nc\geq 1$) and any gauge group.
In the following we shall discuss the case of a specific heterotic vacuum with gauge group $SO(2n)$ at level one. Then we can write the gauge currents $J^a$ as
\eqn\currents{
J^a=\h\ (T^a)_{ij}:\psi^i\psi^j:\ ,\ \ \ a=1,\ldots,n(2n-1)\ ,}
with $2n$ free fermions $\psi^i$ in a real representation of the Lie group $G$. The normalization of the representation matrices $T^a$ satisfies\foot{This normalization corresponds to taking the roots to have length two. An explicit representation of the $SO(2n)$ generators $T^a$ in the vector representation is $(T^{kl})_{ij}=\delta^k_i\delta^l_j-\delta^k_j\delta_i^l$. With this the currents become $J^{kl}=:\psi^k\psi^l:$.}   $\tr(T^aT^b)=2\delta^{ab}$.

Closed string amplitudes decompose into correlators involving only left--movers and
correlators of only right--movers. Hence, the correlator of vertex operators in \TODO\ is evaluated by performing all possible Wick contractions in both sectors separately.

In the scattering amplitude \TODO\ the gauge vertex operators \vertex\ contribute  the current correlator $\vev{J^{a_1}(z_1)\ldots J^{a_N}(z_N)}$ to the left moving sector.
This correlator can be evaluated as a sum over permutations $\rho\in S_N'$ without fixed points \DolanEH
\eqn\Dolan{
\vev{J^{a_1}(z_1)\ldots J^{a_N}(z_N)}=\sum_{\rho\in S_N'\atop \rho=\xi_1\ldots\xi_r}
(-1)^r\ \prod_{l=1}^r f_{\xi_l}\ ,}
with each permutation $\rho$ written as product of cycles, $\rho=\xi_1\ldots\xi_r$ and one cycle
$\xi=(i_1,i_2,\ldots,i_m)$ contributing a factor $f_{\xi}$ with:
\eqn\cycGod{
f_\xi=\h\ \fc{\tr(T^{a_{i_1}}\ldots T^{a_{i_m}})}{z_{i_1i_2}z_{i_2i_3}\ldots z_{i_mi_1}}\ .}
The number $d(N)$ of derangements $S_N'$ is given  by $d(N)=N!\sum\limits_{i=2}^N\fc{(-1)^i}{i!},\ N\geq 2.$ Hence, we have $d(2)=1,\ d(3)=2,\ d(4)=9, d(5)=44,\ldots$. {\it E.g.} for $N=4$ we have the
nine derangements $(2 3 4 1),\ (2 4 1 3),\ (3 1 4 2),\ (3 4 2 1),\ (4 1 2 3),\ (4 3 1 2)$ with $r=1$
and  $(2 1 4 3),\ (3 4 1 2),\ (4 3 2 1)$ with $r=2$, respectively.
In \Dolan\ these nine permutations give rise to \GrossRR:
\eqn\Dolanvier{\eqalign{
\vev{J^{a_1}(z_1)J^{a_2}(z_2)J^{a_3}(z_3) J^{a_4}(z_4)}&=
\fc{1}{4}\ \fc{\tr(T^{a_1}T^{a_2})\ \tr(T^{a_3}T^{a_4})}{z_{12}^2z_{34}^2}-
\fc{\tr(T^{a_1}T^{a_2}T^{a_3}T^{a_4})}{z_{12}z_{23}z_{34}z_{41}}\cr\crr
&-(2\leftrightarrow  3)-(2\leftrightarrow 4)\ .}}
In the following in \Dolan\ we shall be interested in the single trace (s.t.) contributions of \Dolan.
The latter appear for $r=1$, \ie cycles, which do not factorize
into products. Hence the relevant part of \Dolan\ is:
\eqn\Dolani{
\vev{J^{a_1}(z_1)\ldots J^{a_N}(z_N)}_{\rm s.t.}=-\h\
\sum_{\rho\in S_{N-1}} \fc{\tr(T^{a_1}T^{a_{\rho(2)}}\ldots T^{a_{\rho(N)}})}{z_{1\rho(2)}z_{\rho(2),\rho(3)}\ldots z_{\rho(N-1),\rho(N)}z_{\rho(N),1}}\ .}
The correlator \Dolani\ is dressed with the Koba--Nielsen factor
$\prod\limits_{i<j}^{N} |z_{ij}|^{s_{ij}}$ and for a generic group trace
$\tr(T^{a_1}T^{a_{\rho(2)}}\ldots T^{a_{\rho(N)}})$ the total contribution form the left--moving part amounts to
\eqn\leftmovers{
\Ac_L(1,\rho(2,\ldots,N))=-\fc{\prod\limits_{i<j}^{N} |z_{ij}|^{s_{ij}}}{z_{1\rho(2)}z_{\rho(2),\rho(3)}\ldots z_{\rho(N-1),\rho(N)} z_{\rho(N),1}}\ .}

Alternatively, the single trace gauge correlator \Dolani\ may be written as sum over $(N-2)!$ permutations as
\eqn\dolani{
\vev{J^{a_1}(z_1)\ldots J^{a_N}(z_N)}_{\rm s.t}=-2^{N-3}
\sum_{\sigma\in S_{N-2}}\fc{f^{a_1a_{\si(2)}c_1}\ f^{c_1a_{\si(3)}c_2}\ldots f^{c_{N-3}a_{\si(N)}a_{N-1}}}
{z_{1,\si(2)}z_{\si(2),\si(3)}\ldots z_{\si(N-2),\si(N)}z_{\si(N),N-1}z_{N-1,1}},}
with the structure constants $f^{abc}=\h \tr([T^a,T^b]T^c)$. Agreement between the two expressions \Dolani\ and \dolani\ can be shown by using partial fraction
decompositions in the positions $z_i$, $f^{a_1a_2c_1}\ f^{c_1a_3c_2}\ldots f^{c_{N-3}a_N a_{N-1}}=2^{2-N}\tr(T^{a_1}[T^{a_2},[T^{a_3},\ldots,[T^{a_N},T^{a_{N-1}}]\ldots]])$ and the Jacobi relation
$f^{a_1a_2c}f^{ca_3a_4}-f^{a_2a_3c}f^{ca_4a_1}-f^{a_4a_2c}f^{ca_3a_1}=0$.
The form \dolani, which uses a basis of
$(N-2)!$ building blocks  \MafraKJ\ (\cf also \eqq \disk), is suited\foot{Explicitly,
for $N=4$ we have
$\vev{J^{a_1}(z_1)\ldots J^{a_4}(z_4)}_{\rm s.t.}=-2\ \lf(\fc{f^{a_1a_2c}f^{ca_4a_3}}{z_{12}}+\fc{f^{a_1a_4c}f^{ca_2a_3}}{z_{23}}\ri)$, while for $N=5$ we obtain
$\vev{J^{a_1}(z_1)\ldots J^{a_5}(z_5)}_{\rm s.t.}=-4\ \Big(
\fc{f^{a_1a_2c}f^{ca_3d}f^{da_5a_4}}{z_{12}z_{23}}
+\fc{f^{a_1a_3c}f^{ca_2d}f^{da_5a_4}}{z_{13}z_{32}}
+\fc{f^{a_1a_3c}f^{ca_5d}f^{da_2a_4}}{z_{13}z_{24}}
+\fc{f^{a_1a_5c}f^{ca_3d}f^{da_2a_4}}{z_{24}z_{32}}
+\fc{f^{a_1a_2c}f^{ca_5d}f^{da_3a_4}}{z_{12}z_{34}}
+\fc{f^{a_1a_5c}f^{ca_2d}f^{da_3a_4}}{z_{23}z_{34}}\Big)$ for the
gauge choice $z_1=0,z_{N-1}=1,z_N=\infty $.} to extract
BCJ numerators \BCJ\ in terms of structure constants.

In \TODO\ the right--moving sector of the  vertex operators \vertex\ constitutes  the
space--time dependent part of the heterotic amplitude. In fact, this part represents a copy\foot{The right--moving parts of the  vertex operators \vertex\ agree with the open superstring  gluon vertex operators after the replacement $\ap\ra \fc{\ap}{4}$. As a consequence in all our subsequent heterotic results this rescaling still has to be performed.} of the
 open superstring and we may simply borrow the results from the $N$--gluon computation
presented at the beginning of this Section. Hence, before fixing conformal invariance the
right--moving part gives rise to:
\eqnn\rightmovers{
$$\eqalignno{
\Ac_R&=(-1)^{N-3}\ \sum_{\sigma\in S_{N-3}}  A(\sigma)&\rightmovers\cr
&\times\sum_{\ov\rho\in S_{N-3}} \fc{\prod\limits_{i<j}^{N} |\ov z_{ij}|^{s_{ij}}}{\ov z_{1\ov\rho(2)} \ov z_{\ov\rho(2),\ov\rho(3)} \ldots \ov z_{\ov\rho(N-3),\ov\rho(N-2)}\ov z_{\ov\rho(N-2), N}\ov z_{N,N-1}\ z_{N-1,1}}\ S[\ov\rho|\si]\ .}$$}

Therefore, after putting together the pieces from the left--movers \leftmovers\ and right--movers \rightmovers\ and taking into account all normalizations the single trace part of \TODO, which constitutes
\eqn\TODOst{
\Af_{N,\;{\rm s.t.}}^{\rm HET}=(g_{YM}^{\rm HET})^{N-2}\ \sum_{\Pi\in S_{N}/{\bf Z}_2}\tr(T^{a_{\Pi(1)}}\ldots T^{a_{\Pi(N)}})\ \Ac^{\rm HET}(\Pi(1),\ldots,\Pi(N))\ ,}
yields for a generic group trace
$\tr(T^{a_1}T^{a_{\rho(2)}}\ldots T^{a_{\rho(N-2)}}T^{a_{N-1}}T^{a_N})$  the heterotic subamplitude
\eqnn\intresult{
$$\eqalignno{
A^{\rm HET}(\rho)&=V_{\rm CKG}^{-1}\  \lf(\prod_{j=1}^{N}\int\limits_{z_j\in\IC} d^2z_j\ri) \Ac_L(1,\rho(2,\ldots,N-2),N-1,N)\ \Ac_R\cr\crr
&=(-1)^{N-3}\ \sum_{\sigma\in S_{N-3}}\sum_{\ov\rho\in S_{N-3}}
J[\rho\,|\,\ov\rho]\ S[\ov\rho|\si]\ A(\si)\ ,&\intresult}$$}
with the complex integral:
\eqnn\sphereH{
$$\eqalignno{
J[\rho\,|\,\ov\rho]&:=V_{\rm CKG}^{-1}\  \lf(\prod_{j=1}^{N}\int\limits_{z_j\in\IC} d^2z_j\ri)
\ \prod\limits_{i<j}^{N} |z_{ij}|^{2s_{ij}}\ \fc{1}{z_{1\rho(2)}z_{\rho(2),\rho(3)}\ldots z_{\rho(N-2),N-1}z_{N-1,N}z_{N,1}}\cr
&\times\fc{1}{ \ov z_{1\ov\rho(2)} \ov z_{\ov\rho(2),\ov\rho(3)} \ldots \ov z_{\ov\rho(N-3),\ov\rho(N-2)}\ov z_{\ov\rho(N-2), N}\ov z_{N,N-1}\ \ov z_{N-1,1} }\ .&\sphereH}$$}
Note, that in \intresult\ we are concentrating on a basis of $(N-3)!$ heterotic subamplitudes
referring to the color ordering $(1,\rho(2,\ldots,N-3),N-1,N)$. All other heterotic subamplitudes can be expressed in terms of this basis as a result of partial fraction decomposition (\cf \eqq \dolani) and partial integrations in the left--moving sector.
In these lines by using partial fraction decompositions and partial integrations in the left--moving sector we can express the complex integral \sphereH\ as
\eqn\nice{
J[\rho|\ov\rho]=\sum_{\tau\in S_{N-3}} (K^{-1})_{\rho}^{\ \tau}\ I[\tau|\ov \rho]\ ,}
with:
\eqnn\sphere{
$$\eqalignno{
I[\rho\,|\,\ov\rho]&:= \lf(\prod_{j=2}^{N-2}\int\limits_{z_j\in\IC} d^2z_j\ri) \
\prod\limits_{i<j}^{N-1} |z_{ij}|^{2s_{ij}}&\sphere\cr
&\times\fc{1}{  z_{1\rho(2)} z_{\rho(2),\rho(3)} \ldots z_{\rho(N-3),\rho(N-2)}\ \ov z_{1\ov\rho(2)} \ov z_{\ov\rho(2),\ov\rho(3)} \ldots \ov z_{\ov\rho(N-3),\ov\rho(N-2)} }\ .}$$}
In \sphere\ we have chosen the gauge choice  $z_1=0,\ z_{N-1}=1,\ z_N=\infty$ and
the remaining $N-3$ positions $z_i$ are integrated on the whole complex sphere.
The matrix $K_{\rho}^\sigma$ accounts for the basis change of SYM subamplitudes
\eqn\basischange{
\tilde A(\rho)=K_{\rho}^{\ \sigma}\ A(\sigma)}
expressing the basis of $(N-3)!$ SYM subamplitudes
\eqn\subYMi{
\tilde A(\rho):=A(1,\rho(2,\ldots,N-2),N,N-1)}
in terms of the basis \subYM.
According to the arguments given  in \refs{\MafraNVii,\Broedel}, the inverse transposed matrix 
$K^\ast:=(K^{-1})^t$  describes the corresponding basis change on the open string world--sheet integrand of the left--moving sector, \ie $Z_\pi(1,\rho(2,\ldots,N-2),N-1,N)=Z_\pi(1,\tau(2,\ldots,N-2),N,N-1)\ 
(K^\ast)^{\tau}_{\ \rho}$.
{\it E.g.} for $N=4$ we have $K=\fc{s_{23}}{s_{13}}$, while for $N=5$ the matrix:
\eqn\kmatrixv{
K_{\rho}^{\ \sigma}=\pmatrix{\fc{(s_{13}+s_{15})s_{34}}{s_{14}s_{35}}&-\fc{s_{13}s_{24}}{s_{14}s_{35}}\cr\crr
-\fc{s_{12}s_{34}}{s_{14}s_{25}}&\fc{(s_{12}+s_{15})s_{24}}{s_{14}s_{25}}}\ .}
For $N=6$  the first row of $K_{\rho}^{\ \sigma}$ reads
\eqn\kmatrixvi{\eqalign{
K_{1}^{\ \sigma}&=\lf[s_{15}\ s_{46}\ (s_{34}+s_{46}+s_{36})\ri]^{-1}\ \Big\{\ s_{45}\ (s_{13}+s_{14}+s_{16})\ (s_{26}+s_{36}+s_{56}+s_{35}+s_{61}),\cr\crr
&\ \ \ -s_{13}\ s_{25}\ s_{45},\ s_{14}\ s_{25}\ (s_{13}+s_{14}+s_{24}+s_{34}+s_{45}),\
-s_{13}\ s_{25}\ (s_{24}+s_{45}),\cr\crr
&\ \ \ \ -s_{14}\ s_{35}\ (s_{13}+s_{14}+s_{16}),\
s_{35}\ (s_{13}+s_{14}+s_{16})\ (s_{34}+s_{45}+s_{46})\ \Big\}\ ,}}
with the entry $\sigma\in S_{3}$ referring to the permutations $(2,3,4),\ (3,2,4),\ (4,3,2),\
(3,4,2),$\ $(4,2,3)$ and $(2,4,3)$, respectively.
The remaining entries of $K$  may be obtained from \kmatrixvi\
by permuting the numbers $2,3$ and $4$ and changing the positions $\sigma$ in accord with  the basis  $A(\si)$ they refer to.
A general formula for $K$ can be derived by rewriting a similar formula given in \BCJ.
After some adjustments we find (for $N\geq 4$)
\eqn\getformula{
K_{1}^{\ \sigma}=\prod_{l=2}^{N-2} \fc{c(\{1,\sigma(2,\ldots,N-2),N-1\};l)}{(k_N+k_{N-2}+\ldots+k_l)^2}\ ,}
with the functions $c=c_1+c_2$
\eqnn\functionF{
$$\eqalignno{
c_1(\{1,\sigma(2,\ldots,N-2),N-1\};l)&=\cases{
\sum\limits_{o=\tau_l}^{N-1}\kappa_{l,\rho_o}\ , &$\tau_{l+1}<\tau_l\ ,$\cr\crr
-\sum\limits_{o=1}^{\tau_l}\kappa_{l,\rho_o}\ , &$\tau_{l+1}>\tau_l$\ ,}&\functionF\cr\crr\crr
c_2(\{1,\sigma(2,\ldots,N-2),N-1\};l)&=
\cases{(k_N+k_{N-2}+\ldots+k_l)^2\ ,&$\tau_{l+1}<\tau_l<\tau_{l-1}\ ,$\cr\crr
-(k_N+k_{N-2}+\ldots+k_l)^2\ ,&$\tau_{l+1}>\tau_l>\tau_{l-1}\ ,$\cr\crr
0&else\ ,}}$$}
associated to leg $l$. Above $\tau_l$ (with $\tau_1:=0$ and $\tau_{N-1}:=\tau_{N-3}$) is the position of leg $l$ in the set  $\rho:=\{\rho_1,\ldots,\rho_{N-1}\}=\{1,\sigma(2,\ldots,N-2),N-1\}$, and:
\eqn\kap{
\kappa_{ij}=\cases{s_{ij}&$i>j,\ {\rm or}\ j=N-1\ ,$\cr\crr
0&else\ .}}

Finally, with \nice\ we can write \intresult\ in matrix notation as
\eqn\intresulti{
\Ac^{\rm HET}=(-1)^{N-3}\ J\ S\ A=(-1)^{N-3}\ K^{-1}\ I \ S\ A\ ,}
with the vector $\Ac^{\rm HET}$, whose entries $\Ac_{\rho}^{\rm HET}=\Ac^{\rm HET}(\rho)$ are the
$(N-3)!$ heterotic subamplitudes \intresult.
With the  identity  $I=K\ \sv(Z)$, which we will prove in \eqq (4.10), the relations $F=(-1)^{N-3}Z\;S$ and
$\Ac^{\rm I}=F  A$  following from \eqqs \Period\ and \OPEN, respectively we finally have\foot{Note, that with the comments from Footnote 5, the map \mapSV\ has to be accompanied by the rescaling of the inverse string tension $\ap\ra\fc{\ap}{4}$.}:
\eqn\RESULT{
\Ac^{\rm HET}=\sv(\Ac^{\rm I})\ .}
To conclude, the single trace heterotic gauge amplitudes $\Ac^{\rm HET}(\rho)$ referring to the color ordering $\rho$ are simply obtained from the relevant open
string gauge amplitudes $\Ac^{\rm I}(\rho)$ by imposing the projection $\sv$ introduced in \mapSV.
Therefore, the $\ap$--expansion of the heterotic amplitude $\Ac^{\rm HET}(\rho)$ can be obtained from that of the open superstring amplitude $\Ac^{\rm I}(\rho)$
by simply replacing MZVs by their corresponding SVMZVs according to the rules \Example\
introduced in \mapSV.

As corollary let us recall the  relations for the color ordered open superstring subamplitudes
\eqnn\DUAL{
$$\eqalignno{
\Ac^{\rm I}(1,2,\ldots,N)&+e^{i\pi s_{12}}\ \Ac^{\rm I}(2,1,3,\ldots,N-1,N)+
e^{i\pi( s_{12}+ s_{13})}\ \Ac^{\rm I}(2,3,1,\ldots,N-1,N)\cr
&+\ldots+e^{i\pi( s_{12}+ s_{13}+\ldots+ s_{1,N-1})}\ \Ac^{\rm I}(2,3,\ldots,N-1,1,N)=0\ ,&\DUAL}$$}
and permutations thereof following from considering monodromies on the open string world--sheet
\refs{\StiebergerHQ,\BjerrumBohrRD}. Applying the map \mapSV\ on \DUAL, thereby using $\sv(\pi)=0$ and applying \RESULT\ gives the set of KK equations
\eqnn\DUALi{
$$\eqalignno{
\Ac^{\rm HET}(1,2,\ldots,N)&+ \Ac^{\rm HET}(2,1,3,\ldots,N-1,N)+
 \Ac^{\rm HET}(2,3,1,\ldots,N-1,N)\cr
&+\ldots+ \Ac^{\rm HET}(2,3,\ldots,N-1,1,N)=0\ ,&\DUALi}$$}
and the BCJ relations
\eqnn\DUALii{
$$\eqalignno{
s_{12}\ &\Ac^{\rm HET}(2,1,3,\ldots,N-1,N)+(s_{12}+ s_{13})\ \Ac^{\rm HET}(2,3,1,\ldots,N-1,N)\cr
&+\ldots+( s_{12}+ s_{13}+\ldots+ s_{1,N-1})\ \Ac^{\rm HET}(2,3,\ldots,N-1,1,N)=0\ ,&\DUALii}$$}
for the heterotic subamplitudes, respectively in agreement with our comments after \eqq \sphereH.
To conclude, heterotic string subamplitudes fulfil the same
amplitude relations, \ie KK and BCJ relations as field--theory amplitudes.

\newsec{Representation of gravitational amplitudes in superstring theory}

In this Section we shall elaborate on the graviton tree--level scattering amplitude in superstring theory and relate it to the heterotic gauge amplitude computed in the previous Section.
Furthermore, we will discuss type II and heterotic scalar amplitudes and related them to type I
scalar amplitudes.

Thanks to the KLT relations \KawaiXQ\ at tree--level closed string amplitudes can be expressed as sum over squares of
(color ordered)  open string subamplitudes arising from the left-- and right--moving sectors.
This map gives a relation between a closed string tree--level amplitude involving $N$ closed strings and a sum of squares of (partial ordered) open string tree--level amplitudes each  involving $N$ open strings.
{\it E.g.} for
the $N$--graviton scattering amplitude $\Mc$ in (type I or type II) superstring theory we may write these identities as follows \refs{\KawaiXQ,\BernSV,\Bohr}
\eqnn\Graviton{
$$\eqalignno{
\Mc(1,\ldots,N) &= (-1)^{N-3} \ \kappa^{N-2}\ \sum_{\si \in S_{N-3}} \Ac(1,\si(2,3,\ldots,N-2),N-1,N)  \cr
&\times \ \sum_{\rho \in S_{N-3}} \Sc[\rho|\si] \ \tilde \Ac(1,\rho(2,3,\ldots,N-2),N,N-1)\ ,&\Graviton}$$}
with the gravitational coupling constant $\kappa$ and  a product of $\sin$--factors  
\eqn\Kernel{
\Sc[\rho|\si]:=\Sc[\, \rho(2,\ldots,N-2) \, | \, \si(2,\ldots,N-2) \, ]= \prod_{j=2}^{N-2} \sin\Big( \, s_{1,j_\rho} \ + \ \sum_{k=2}^{j-1} \theta(j_\rho,k_\rho) \, s_{j_\rho,k_\rho} \, \Big)\ ,}
which depend on the kinematic invariants \Mandel\ and arise from the KLT relations \refs{\KawaiXQ,\Bohr}. In \Kernel\ we use the  the same notation as described below \eqq \kernel.
In \Graviton\ the graviton amplitude is expressed as a sum over $[(N-3)!]^2$ terms each contributing a product of two full--fledged open superstring amplitudes $\Ac$ and $\tilde \Ac$.
The expression \Graviton\ is a very intricate way of writing a closed string amplitude  $\Mc$. In fact,
 in \StiebergerWEA\ it has already been anticipated, that there are more efficient and elegant ways in  writing \Graviton\ in terms of a sum involving linearly $(N-3)!$ single full--fledged open superstring
 amplitudes $\Ac$ only.

Building up on the open superstring result \OPEN\ the $N$--graviton amplitude in superstring theory can be written as
\eqn\KLT{
\Mc(1,\ldots,N)=  \kappa^{N-2}\ \sum_{\si \in S_{N-3}} \sum_{\ov\si \in S_{N-3}}
  A(\si)\ G[\si | \ov\si]\ A(\ov\si)\ ,}
with the gravity kernel
\eqnn\GKernel{
$$\eqalignno{
G[\si|\ov\si]&:=G[\si(2,3,\ldots,N-2)\, | \, \ov\si(2,3,\ldots,N-2)]\cr\crr
&=\sum_{\rho \in S_{N-3}}\sum_{\ov\rho \in S_{N-3}}
S[\rho|\si]\ I[\rho\,|\,\ov\rho]\  S[\ov\rho| \ov\si]\ ,&\GKernel}$$}
and the world--sheet sphere integrals introduced in \eqq \sphere.
The field--theory limit of the graviton amplitude \KLT\ can be written as
\eqn\limitG{
\Mc_{FT}(1,\ldots,N)= (-1)^{N-3} \ \kappa^{N-2}\ \sum_{\si \in S_{N-3}} \sum_{\ov\si \in S_{N-3}}
  A(\si)\ S_0[\si | \ov\si]\ A(\ov\si)\ ,}
with the intersection matrix $S_0$ whose entries are $S_{0\;\rho,\si}:=S_0[\rho|\si]$.
The limit \limitG\ has to agree with the expression (with $\tilde A$ defined in \subYMi)
\eqn\limitGa{
\Mc_{FT}(1,\ldots,N)= (-1)^{N-3} \ \kappa^{N-2}\ \sum_{\si \in S_{N-3}} \sum_{\ov\si \in S_{N-3}} A(\si)\
 S[ \si|\ov \si ] \  \tilde A(\ov \si)\ ,}
which directly follows from \Graviton. Comparing the two expressions \limitG\ and \limitGa\ and using the basis change \basischange\ yields:
\eqn\yields{
S_0=S\ K\ .}

One key observation in \StiebergerWEA\ is that the closed  superstring amplitude
\KLT\ can be expressed in terms of a linear combination of $(N-3)!$ open superstring amplitudes \OPEN\ as
\eqn\KEY{
\Mc=(-1)^{N-3}\ \kappa^{N-2}\ A^t\ S_0\ \sv(\Ac^{\rm I})\ ,}
subject to the map $\sv$ introduced in \mapSV\ and the intersection matrix $S_0$ describing the field--theory limit \limitG\ of the graviton amplitude \KLT.
As a consequence of \KEY\ for the gravity kernel \GKernel\ we have
\eqn\Key{
G=S^tIS=(-1)^{N-3}\ S_0\ \sv(F)\ ,}
\ie with $F=(-1)^{N-3}ZS$ from \Period\ and \yields\ we obtain:
\eqn\followI{
I=K\ \sv(Z)\ .}
With our result \RESULT\ we can write the tree--level graviton $N$--point amplitude \KEY\ of superstring theory as
\eqn\RESULTT{
\Mc=(-1)^{N-3}\ \kappa^{N-2}\ A^t\ S_0\ \Ac^{\rm HET}\ ,}
with the vector $\Ac^{\rm HET}$ of $(N-3)!$ heterotic single trace tree--level gauge $N$--point  amplitudes \RESULT.
Note, that by applying naively KLT relations we would not have arrived at \RESULTT.
The relation \RESULTT\ connects two seemingly different amplitudes from two
different string vacua.
This might provide an extension to the heterotic--type I duality \PolchinskiDF, see also Section~6.

Obviously, in four--dimensional $\Nc=8$ type II superstring vacua the relationship  \KEY\ 
can be applied for closed string amplitudes involving scalars. 
The latter belong to the supergravity multiplet and a subclass $\Phi^{ij},\ i,j=1,\ldots,6$ of them denote geometric moduli fields describing the internal $D=6$ toroidal compactification with the metric $g^{ij}$.
For instance, the four--scalar amplitude in four--dimensional $\Nc=8$  type II string vacua reads
\eqn\ScalarII{\eqalign{
\Ac^{\rm II}(\Phi_1^{i_1j_1},\Phi_2^{i_2j_2},\Phi_3^{i_3j_3},\Phi_4^{i_4j_4}) &=\fc{u}{st}\
\lf(t\ \delta_1+s\ \delta_2+\fc{st}{u}\ \delta_3\ \ri)\cr\crr
&\times\lf(t\ \ov\delta_1+s\ \ov\delta_2+\fc{st}{u}\ \ov\delta_3\ \ri)\
\fc{\Gamma(s)\ \Gamma(u)\ \Gamma(t)}{\Gamma(-s)\ \Gamma(-u)\ \Gamma(-t)}\ ,}}
with:
\eqnn\Deltas{
$$\eqalignno{
\delta_1&=g^{i_1i_2}\ g^{i_3i_4}\ \ \ ,\ \ \ \delta_2=g^{i_1i_3}\ g^{i_2i_4}\ \ \ ,\ \ \ \delta_3=g^{i_1i_4}\ g^{i_2i_3},\ \cr
\ov\delta_1&=g^{j_1j_2}\ g^{j_3j_4}\ \ \ ,\ \ \ \ov\delta_2=g^{j_1j_3}\ g^{j_2j_4},\ \ \ ,\ \ \
\ov\delta_3=g^{j_1j_4}\ g^{j_2j_3}\ .&\Deltas}$$}
The kinematic invariants \Mandel\ are given by $s=\ap(k_1+k_2)^2,\ t=\ap(k_1+k_3)^2,\ 
u=\ap(k_1+k_4)^2$, with $s+t+u=0$. On the other hand, in four--dimensional $\Nc=4$  type I  string vacua the four--point subamplitude
 of scalar fields $\Phi^j,\ j=1,\ldots,6$ describing  open string moduli (transversal $D$--brane positions or Wilson lines) reads
\eqn\ScalarI{
\Ac^{\rm I}(\Phi_1^{j_1},\Phi_2^{j_2},\Phi_3^{j_3},\Phi_4^{j_4}) =\lf(\ t\ \ov\delta_1+s\ \ov\delta_2+\fc{st}{u}\ \ov\delta_3\ \ri)\ \fc{\Gamma(s)\ \Gamma(1+u)}{\Gamma(1+s+u)}\ ,}
w.r.t. to the color ordering $(1,2,3,4)$ and the symbols defined in \Deltas.
Comparing the two amplitudes \ScalarII\ and \ScalarI\ and using \eqq (5.3) yields the correspondence 
\eqn\ScalarIII{\eqalign{
\Ac^{\rm II}(\Phi_1^{i_1j_1},\Phi_2^{i_2j_2},\Phi_3^{i_3j_3},\Phi_4^{i_4j_4}) &=
-\fc{u}{t}\ \lf(t\ \delta_1+s\ \delta_2+\fc{st}{u}\ \delta_3\ \ri)\cr\crr
&\times \sv\lf( \Ac^{\rm I}(\Phi_1^{j_1},\Phi_2^{j_2},\Phi_3^{j_3},\Phi_4^{j_4}) \ri)\ ,}}
in lines of \KEY. Similar relation  than \ScalarIII\ can also be derived for amplitudes involving more than four scalar fields.

Finally, let us briefly discuss the heterotic analog of the scalar amplitude \ScalarII. The four--scalar amplitude in four--dimensional $\Nc=4$  heterotic string vacua reads
\eqn\ScalarHET{\eqalign{
\Ac^{\rm HET}(\Phi_1^{i_1j_1},\Phi_2^{i_2j_2},\Phi_3^{i_3j_3},\Phi_4^{i_4j_4}) &=\fc{u}{st}\ \lf(\ \fc{t}{1-s}\
\ \delta_1+\fc{s}{1-t}\ \delta_2+\fc{t}{1-u}\ \fc{s}{u}\ \delta_3\ \ri)\cr\crr
&\times\lf(t\ \ov\delta_1+s\ \ov\delta_2+\fc{st}{u}\ \ov\delta_3\ \ri)\
\fc{\Gamma(s)\ \Gamma(u)\ \Gamma(t)}{\Gamma(-s)\ \Gamma(-u)\ \Gamma(-t)}\ ,}}
with \Deltas.
An expression similar to \ScalarHET\ can be derived for the four--scalar amplitude involving scalars, which describe Wilson line moduli. In four--dimensional $\Nc=4$ heterotic string theory the scalars 
$\Phi^{ij}$  belong to vector multiplets. 
In \ScalarHET\ the spurious tachyonic poles, which arise from the exchange of fields from the massless $\Nc=4$ supergravity multiplet, have no counterpart in the tree--level perturbative type I superstring amplitude \ScalarI. Nevertheless, the full $\ap$--dependence  of the heterotic amplitude \ScalarHET\ can again be described by the open string amplitude \ScalarI.
Comparing the two amplitudes \ScalarHET\ and \ScalarI\ and using \eqq (5.3) yields:
\eqn\ScalarHETI{\eqalign{
\Ac^{\rm HET}(\Phi_1^{i_1j_1},\Phi_2^{i_2j_2},\Phi_3^{i_3j_3},\Phi_4^{i_4j_4}) &=
-\fc{u}{t}\ \lf(\ \fc{t}{1-s}\
\ \delta_1+\fc{s}{1-t}\ \delta_2+\fc{t}{1-u}\ \fc{s}{u}\ \delta_3\ \ri)\cr\crr
&\times \sv\lf( \Ac^{\rm I}(\Phi_1^{j_1},\Phi_2^{j_2},\Phi_3^{j_3},\Phi_4^{j_4}) \ri)\ .}}
Amplitudes involving more than four scalar fields can be cast into a similar form than \ScalarHETI.
Note, that by applying naively KLT relations we would not have arrived at  \ScalarHETI.
Connections in lines of  \ScalarHETI\ can also be established for heterotic gravitational amplitudes or non--single trace  gauge amplitudes subject to gravitational exchanges.

The lesson to learn from the example \ScalarHETI\ is that also closed string  amplitudes other than heterotic (single--trace) gauge \RESULT\ or
superstring gravitational amplitudes \KEY\ can be expressed as single--valued image of some open string amplitudes. 
Generically, after performing partial fraction decompositions and partial integration relations any complex integral referring to a closed string world--sheet integral can be expressed in terms of the fundamental basis \sphereH, which serves as building block for complex integrals.
The elements of this basis can be written as single--valued image of some open string world--sheet integrals (\cf Section 5). Therefore, any closed string amplitude can be written as single--valued image of  open string amplitudes by expressing the underlying closed string world--sheet integrals as
single--valued image of open string integrals.
As a consequence the whole $\ap$--dependence of closed string amplitudes is entirely encapsulated in the  corresponding open string amplitude.

\newsec{Complex  vs. iterated integrals: closed vs. open string world--sheet integrals}

Open string world--sheet disk integrals \disk\ are described as iterated (real) integrals on
$\IR\IP^1\backslash\{0,1,\infty\}$, while closed string world--sheet sphere integrals \sphereH\ are
given by integrals over the full complex plane. The latter, which can be considered as iterated integrals on $\IP^1\backslash\{0,1,\infty\}$ integrated independently on all choices of paths,
are more involved than the real iterated integrals appearing in open string amplitudes.
Nevertheless, in the previous two sections we have exhibited non--trivial relations between open and closed string amplitudes and in this Section we shall elaborate on these connections at the level of the world--sheet integrals.
In this Section we shall show, that quite generally complex integrals can be expressed as real iterated integrals subject to the projection $\sv$.

Recall, that \eqq \followI\ expresses complex sphere integrals \sphere\ in terms
of a linear combination of disk integrals \disk\ subject to the map $\sv$.
This is to be contrasted with the KLT formula \Graviton, where squares of
disk integrals \disk\ appear.
In light of \followI\ let us discuss the simplest case describing the scattering of four closed strings. For $N=4$ the real integral \disk\ becomes
\eqn\realvier{
Z:=Z_1(1)=-\int_0^1 dx\ x^{s-1}\ (1-x)^u=-\fc{\Gamma(s)\ \Gamma(1+u)}
{\Gamma(1+s+u)}\ ,}
while the complex integral \sphere\ boils down to:
\eqn\consideration{
I:=I[1|1]=\int_\IC d^2z\ |z|^{2s-2}\ |1-z|^{2u}=\fc{u}{st}\ \fc{\Gamma(s)\ \Gamma(u)\ \Gamma(t)}{\Gamma(-s)\ \Gamma(-u)\ \Gamma(-t)}\ .}
With $K=\fc{u}{t}$ \eqq \followI\  gives rise to:
\eqn\nontrivialone{
\fc{u}{st}\ \fc{\Gamma(s)\ \Gamma(u)\ \Gamma(t)}{\Gamma(-s)\ \Gamma(-u)\ \Gamma(-t)}=-\fc{u}{t}\ \sv\lf(\fc{\Gamma(s)\ \Gamma(1+u)}{\Gamma(1+s+u)}\ri)\ ,}
\ie:
\eqn\nontrivialtwo{
\int_\IC d^2z\ |z|^{2s-2}\ |1-z|^{2u}=-\fc{u}{t}\
\sv\lf(\int_0^1 dx\ x^{s-1}\ (1-x)^u\ri)\ .}
Similar explicit correspondences \followI\ between
complex sphere integrals $I$ and real disk integrals $Z$
can be made for $N\geq 5$. Finally, let us note, that with \wehave\ the complex integrals \sphere\
have the following field--theory limit:
\eqn\Wehavee{
\lf.I\ri|_{\ap=0}=(-1)^{N-3}\ K\ S^{-1}\ .}

Moreover, a direct correspondence between complex sphere integrals and real disk integrals can be made for the (heterotic) world--sheet integrals \sphereH. Indeed, with \nice, \ie $I=K\ J$ \eqq \followI\ becomes:
\eqn\NICE{
J=\sv(Z)\ .}
As one implication of \NICE\ to each {\it single} complex sphere integral $J[\pi|\rho]$ {\it one}
real integral $Z_\pi(\rho)$ corresponds.
{}For our $N=4$ example we now have:
\eqn\consideration{
J:=J[1|1]=-\int_\IC d^2z\ |z|^{2s-2}\ |1-z|^{2u}\ (1-z)^{-1}=\fc{1}{s}\ \fc{\Gamma(s)\ \Gamma(u)\ \Gamma(t)}{\Gamma(-s)\ \Gamma(-u)\ \Gamma(-t)}\ .}
With \realvier\ \eqq \NICE\ gives rise to:
\eqn\Nontrivialtwo{
\int_\IC d^2z\ \fc{|z|^{2s}\ |1-z|^{2u}}{z\ (1-z)\ \ov z}=
\sv\lf(\int_0^1 dx\ x^{s-1}\ (1-x)^u\ri)\ .}
Hence, in \sphereH\ the effect of inserting the left--moving gauge part \leftmovers\ is simply the projection \mapSV\ acting on  the right--moving part \disk.
Similar explicit and direct correspondences \NICE\ between the
complex sphere integrals $Z$ and the real disk integrals $Z$ can be made for $N\geq 5$.
In order to familiarize with the matrix notation let us explicitly write the case \NICE\
for $N=5$ (with $z_1=0,\ z_4=1$):
\eqnn\NICEfive{
$$\eqalignno{
&\hskip-0.75cm\pmatrix{
\displaystyle{\int\limits_{z_2,z_3\in\IC}d^2z_2\ d^2z_3\ \fc{\prod\limits_{i<j}^4|z_{ij}|^{2s_{ij}}}{z_{12}z_{23}z_{34}\ \ov z_{12}\ov z_{23}}}&
\displaystyle{\int\limits_{z_2,z_3\in\IC}d^2z_2\ d^2z_3\ \fc{\prod\limits_{i<j}^4|z_{ij}|^{2s_{ij}}}{z_{12}z_{23}z_{34}\ \ov z_{13}\ov z_{32}}}\cr\crr
\displaystyle{\int\limits_{z_2,z_3\in\IC}d^2z_2\ d^2z_3\ \fc{\prod\limits_{i<j}^4|z_{ij}|^{2s_{ij}}}{z_{13}z_{32}z_{24}\ \ov z_{12}\ov z_{23}}}&
\displaystyle{\int\limits_{z_2,z_3\in\IC}d^2z_2\ d^2z_3\ \fc{\prod\limits_{i<j}^4|z_{ij}|^{2s_{ij}}}{z_{13}z_{32}z_{24}\ \ov z_{13}\ov z_{32}}}}\cr\crr\crr
&\hskip1cm=
\sv\pmatrix{
\displaystyle{\int\limits_{0<z_2<z_3<1}dz_2\ dz_3\ \fc{\prod\limits_{i<j}^4|z_{ij}|^{s_{ij}}}{z_{12}z_{23}}}&
\displaystyle{\int\limits_{0<z_2<z_3<1}dz_2\ dz_3\ \fc{\prod\limits_{i<j}^4|z_{ij}|^{s_{ij}}}{z_{13}z_{32}}}\cr\crr
\displaystyle{\int\limits_{0<z_3<z_2<1}dz_2\ dz_3\ \fc{\prod\limits_{i<j}^4|z_{ij}|^{s_{ij}}}{z_{12}z_{23}}}&
\displaystyle{\int\limits_{0<z_3<z_2<1}dz_2\ dz_3\ \fc{\prod\limits_{i<j}^4|z_{ij}|^{s_{ij}}}{z_{13}z_{32}}}}\ .&
\NICEfive}$$}
In \NICEfive\ we explicitly see how the presence of the left--moving gauge insertion in the complex integrals results in the projection onto real integrals involving only the right--moving part.
Besides, let us compute the closed string analog of \wehave. With \NICE\ and \wehave\ we find:
\eqn\Wehave{
\lf.J\ri|_{\ap=0}=(-1)^{N-3}\ S^{-1}\ .}
Hence, the set of complex world--sheet sphere integrals \sphereH\ are the closed string analogs of the open string world--sheet disk integrals \disk.

To conclude, after applying partial integrations to remove double poles, which are responsible for 
spurious tachyonic poles, all closed superstring
amplitudes can be expressed in terms of the basis \sphereH, which in turn through \NICE\ can be
related to the basis of open string amplitudes \disk. As a consequence the $\ap$--dependence
of any closed string amplitude is given by that of the underlying open string amplitudes, \cf
\eqqs \RESULT, \KEY\ and \ScalarHETI\ as some examples.

Finally, we would like to make a connection to  {\it Ref.} \CachazoIEA. In this reference it has been argued, that the field--theory limit
\wehave\ of the open string world--sheet disk integrals \disk\  is related to the double partial amplitudes of a massless colored cubic scalar theory
\eqn\doublepartial{
m_N^{(0)}(\al|\bet)= \int \fc{d\,^n\sigma}{\rm{Vol}(SL(2,\IC))}\ \fc{\prod\limits_a{}'\delta(\sum\limits_{b\neq a}\ \fc{s_{ab}}{\sigma_{a b}})}{(\sigma_{\alpha(1),\alpha(2)}\cdots\sigma_{\alpha(N),\alpha(1)})\
(\sigma_{\beta(1),\beta(2)}\cdots\sigma_{\beta(N),\beta(1)})}\ ,}
evaluated at the solutions of the scattering equations
$\sum\limits_{b\neq a} \fc{s_{ab}}{\sigma_{a b}}=0$.
More precisely, with $(m_{\rm scalar})_{\al\beta}:=m_N^{(0)}(1,\al(2,\ldots,N-2),N-1,N|1,\bet(2,\ldots,N-2),N,N-1)$ we have \CachazoIEA:
\eqn\Doublepartial{
(m_{\rm scalar})_{\al\beta}=\lf.Z\ri|_{\ap=0}=(-1)^{N-3}\ (S^{-1})_{\al\bet}\ .}
With \Wehave\ we now also obtain a relation to the building blocks \sphereH\  of the heterotic string amplitudes as:
\eqn\CHE{
(m_{\rm scalar})_{\al\beta}=\lf.J\ri|_{\ap=0}=(-1)^{N-3}\ (S^{-1})_{\al\bet}\ .}
It is quite striking, that the structure of \doublepartial\ and \Doublepartial, furnished by the permutations $\al$ and $\beta$, is captured by the left-- and right--moving
parts in \sphereH, respectively. Recently in {\it Ref.} \GomezWZA, a similar observation has been made in the heterotic version of Berkovits new twistor--like superstring theory \BerkovitsXBA.
It would be very interesting to find further connections between the work \CachazoIEA\ (and also \GomezWZA) and perturbative heterotic string amplitudes presented here.

\newsec{Unity of tree--level superstring amplitudes}

It is well known that various formulations/compactifications of superstring theory are connected by a web of dualities. They can be interpreted as different vacua of a universal M--theory \WittenEX. The classic example is type IIA/K3 -- heterotic/T$^4$ duality in six dimensions \HullYS. Similar to many other examples, it is a strong--weak coupling duality. Perturbative states on one side, like heterotic gauge bosons, are mapped to non--perturbative states on the other side, like D--branes wrapping on K3 cycles. There are convincing arguments, in all known duality cases, that such correspondence holds at the massless level. It is not clear, however, what is the r\^ole of Regge excitations in strong--weak coupling dualities.
It is regrettable because, without exaggerating,  the Regge states, as arising from string vibrations, are the true essence of string theory. The $\alpha'$--dependence of the amplitudes discussed in this paper are
due to such Regge states propagating in all possible channels.

One notable exception is type I -- heterotic duality in four dimensions \PolchinskiDF. There is a class of effective action terms, essentially describing the (generalized) non--Abelian Born--Infeld action,
which appears at the
tree--level on type I side, while it is  generated by loop corrections on the heterotic side \TseytlinFY. Massive string excitations appear to play some r\^ole in this correspondence because on  type I side, the Born--Infeld terms appear in the $\alpha'$--loop expansion of the two--dimensional world--sheet sigma model, while in space--time, the corresponding interactions are mediated by Regge states at the tree level. The comparison with heterotic theory works well at the one--loop level 
${\cal O}(F^4)$, but  runs into problems at two loops~${\cal O}(F^6)$~\StiebergerFH.

The single--valued projection connecting type I and heterotic single--trace amplitudes creates a new link in the web of relations shown in Figure 1. Although similar to the duality web, the  ``web of amplitudes'' links the scattering amplitudes evaluated (to all orders in~$\ap$) not only in different string vacua
but, what is most important, it also includes some links between the amplitudes involving external particles {\it not} related by supersymmetry or any known symmetry, like gravitons and gauge bosons.
\ifig\unity{Unity of tree--level superstring amplitudes.}
{\epsfxsize=0.6\hsize\epsfbox{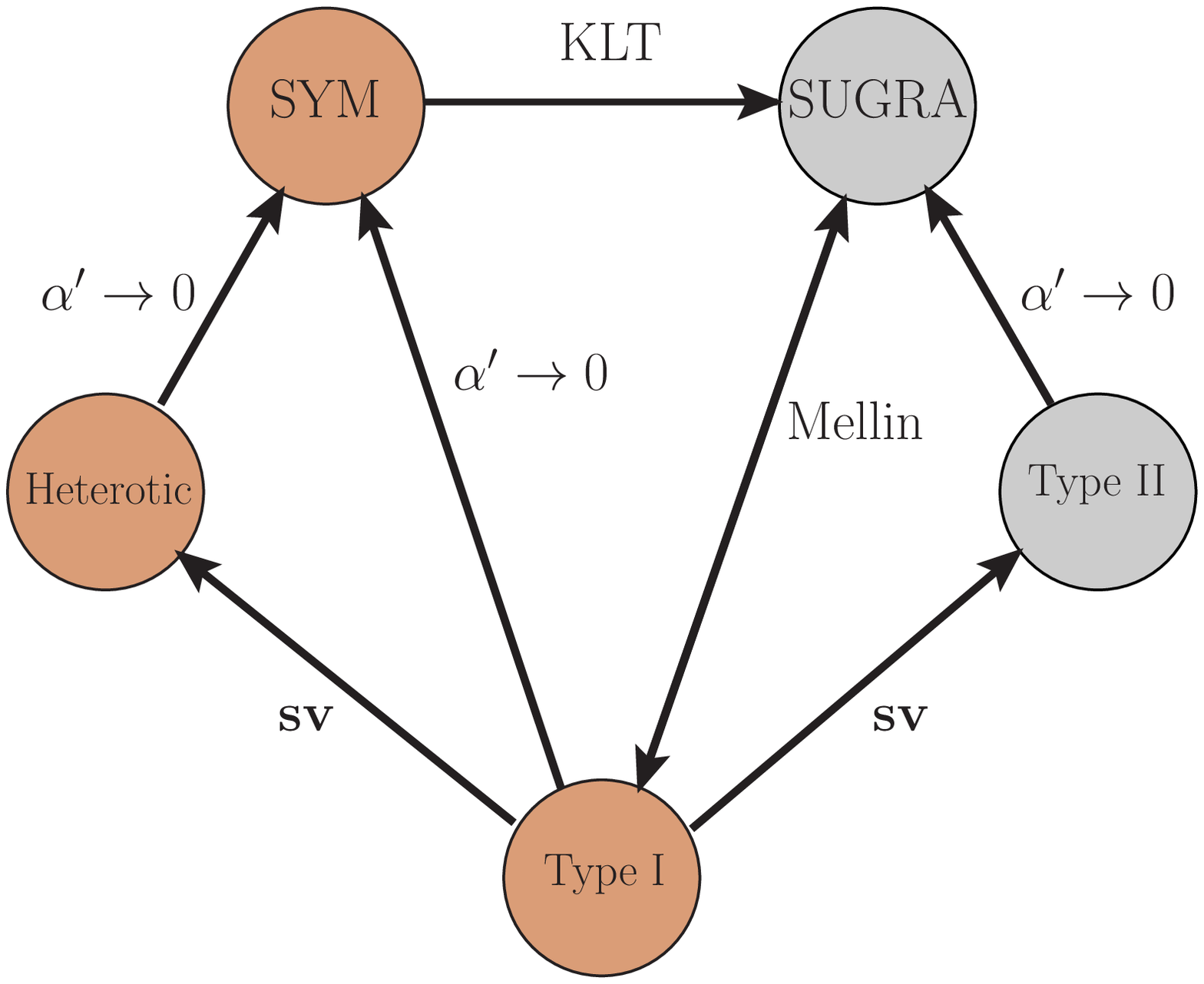}}
\noindent

Type I open string theory appears to play a central r\^ole in the web of amplitudes. By single--valued projections, it generates single--trace heterotic gauge amplitudes \RESULT\ and type II graviton amplitudes \KEY\ \StiebergerWEA. {\it Eq.} \RESULTT\ provides though a bridge from  gauge heterotic to type II graviton amplitude bypassing type I, without using sv projections. Type I connects also, via Mellin correspondence, to $\Nc=8$ supergravity \StiebergerHZA. On the other hand, KLT allows constructing supergravity amplitudes from $\Nc=4$ SYM which supplies the basic building blocks for all tree--level open and closed string amplitudes.

It should be made clear that the connections depicted in Figure 1 appear in perturbation theory, at
the tree--level, in four space--time dimensions. The form of vertex operators is determined by
world--sheet supersymmetry, but their world--sheet correlators are decoupled from the internal sector of SCFT associated to compact dimensions. In order to see the effects of internal dimensions, one would have to consider the amplitudes involving moduli fields like in \eqq \ScalarHETI; one could also go beyond the tree--level to observe internal states and their Regge excitations propagating in  the loops. One of the nodes obviously missing in Figure~1 are the heterotic amplitudes involving external gravitons, which must be sensitive to the massive string spectrum with a lower, $\Nc=4$ supersymmetry, as compared to $\Nc=8$ of type II, \cf also the comments at the end of Section 4.

There must be a deep reason for the universal $\alpha'$ dependence of all tree--level string amplitudes and their connection to SYM and supergravity. It is possible that some new insights can be gained by connecting the web of amplitudes with string dualities.

\newsec{Concluding remarks}

In this work we have found a correspondence between closed and open superstring amplitudes communicated by the sv map \mapSV.
This map relates two string amplitudes of different world--sheet topologies.
One basic example is the relation \RESULT\ between the single trace heterotic tree--level gauge amplitudes  \TODOst\ and open superstring tree--level gauge amplitudes \DECO. Based on this example many other closed/open amplitude connections can be established, \eg relations \ScalarIII\ or \ScalarHETI\ between type II or heterotic and type I scalar amplitudes, respectively.
The essential property common to all such relationships is that the full $\ap$--dependence of the type II or heterotic closed string amplitudes is encapsulated in the type I open string amplitudes.
These relations give rise to a much deeper connection between open and closed string amplitudes than what is implied by KLT relations.
It would be interesting to understand  the sv map \mapSV\ in the framework of sigma--model expansion in the underlying superconformal world--sheet theory. 
Also important is to clarify the r\^ole of the map \mapSV\  at the level of perturbation theory of open and closed strings from the nature  of their underlying string world--sheets.

Furthermore, we have established a connection between the single trace heterotic tree--level gauge amplitudes and  graviton amplitudes of superstring theory, \cf \RESULTT.
This result is quite surprising because it relates gauge and gravitational amplitudes in two different string vacua. On the other hand, in four space--time dimensions one has the relation  \refs{\BershadskyTA,\KaplunovskyJW}
\eqn\KL{
\Delta_{E_6}-\Delta_{E_8'}=12\ F_1\ ,}
which connects type II one--loop superstring corrections to $R^2$ expressed by the topological one--loop partition function $F_1$ (which in turn is related to the  generalized  $\Nc=2$ index) to a difference of one--loop gauge corrections of heterotic $(2,2)$
Calabi--Yau vacua. In {\it Ref.} \KaplunovskyJW\ the
relation \KL\ is explained at the level of the underlying world--sheet superconformal field theory
 as a consequence of the bosonic/supersymmetric map \LSW. It is possible that the sv map is connected to such  a bosonic/supersymmetric map.

The growing set of interconnections hints towards a fascinating unity of closed and open string amplitudes with gauge theory and supergravity.

\vskip2.5cm
\goodbreak
\leftline{\noindent{\bf Acknowledgments}}

\noindent
We gratefully acknowledge support from the Simons Center for Geometry and Physics, Stony Brook University at which a substantial portion of the research for this work was performed.
This material is based in part upon work supported by the National Science Foundation under Grants No.\ PHY-0757959 and PHY-1314774.   Any
opinions, findings, and conclusions or recommendations expressed in
this material are those of the authors and do not necessarily reflect
the views of the National Science Foundation.

\listrefs

\end